\documentclass[10pt]{article} 
\usepackage[preprint]{tmlr}

\usepackage{amsmath,amsfonts,bm}

\def\eqref#1{equation~\ref{#1}}

\def\1{\bm{1}}

\DeclareMathAlphabet{\mathsfit}{\encodingdefault}{\sfdefault}{m}{sl}
\SetMathAlphabet{\mathsfit}{bold}{\encodingdefault}{\sfdefault}{bx}{n}

\usepackage{hyperref}
\usepackage{url}
\usepackage{graphicx}

\usepackage{algorithm}
\usepackage{pgfplots}
\usetikzlibrary{intersections}
\usepgfplotslibrary{fillbetween}
\usepackage{booktabs} 
\usepackage{tabularx,array}
\newcolumntype{L}{>{\raggedright\arraybackslash}X}
\renewcommand{\arraystretch}{1.12}
\usepackage{xcolor}
\renewcommand{\arraystretch}{1.12}
\usepackage{tikz}
\usetikzlibrary{arrows.meta,positioning,shapes.geometric,calc}
\usetikzlibrary{matrix}
\usepackage{amsmath}
\usepackage{caption}
\usepackage{subcaption}
\usepackage{multirow}
\usepackage{amsfonts}
\usepackage{fontawesome5}
\usepackage{minted}
\usepackage[ruled,vlined,noline,algo2e]{algorithm2e}
\usepackage{float} 
\setminted{
  fontsize=\footnotesize,
  linenos=false,
  breaklines=true,
  frame=single,
  bgcolor=white
}
\usepackage{makecell}
\usetikzlibrary{matrix,fit,backgrounds,shapes.symbols}
\definecolor{colA}{RGB}{255,87,51}   
\definecolor{colS}{RGB}{72,149,239}  
\definecolor{colT}{RGB}{88,177,159}  
\definecolor{colG}{RGB}{140,120,200} 
\definecolor{colV}{RGB}{80,80,80}    
\usetikzlibrary{decorations.markings}

\newcommand{\MARKER}{\texttt{\detokenize{<MARKER>}}} 

\newcommand{\NORMW}{\textsf{normalize}(W)}
\usepackage{hyperref}

\newcommand{\upar}[1]{\textsubscript{$\uparrow(#1)$}}
\newcommand{\downar}[1]{\textsubscript{$\downarrow(#1)$}}
\newcommand{\uppar}[1]{\textsubscript{$\uparrow$}}
\usepackage{colortbl}
\usepackage{xcolor}
\definecolor{lightgray}{gray}{0.93}
 \usetikzlibrary{patterns}
\pgfplotsset{compat=1.17}
\usepgfplotslibrary{groupplots}
\usepackage{pgfplotstable}
\usepackage{filecontents}
\pgfdeclarelayer{bg}
\pgfsetlayers{bg,main, background}
\newcolumntype{C}{>{\centering\arraybackslash}X}

\title{Conjunctive Poisoning in AI Supply-Chain Applications}

\author{\name Nokimul Hasan Arif \email no643252@ucf.edu \\
      \addr University of Central Florida
      \AND
      \name Qian Lou \email qian.lou@ucf.edu \\
      \addr University of Central Florida
      \AND
      \name Mengxin Zheng \email mengxin.zheng@ucf.edu \\
      \addr University of Central Florida}

\def\month{MM}  
\def\year{YYYY} 
\def\openreview{\url{https://openreview.net/forum?id=XXXX}} 

\begin{document}

\maketitle

\begin{abstract}
Large Language and Vision-Language Models are increasingly deployed through inference pipelines that include prompt wrappers (e.g., templates and post-processing scripts) and configuration metadata (e.g., JSON/YAML files) that together shape model outputs. While model weights and binaries are routinely verified, these textual deployment artifacts remain weakly protected despite directly influencing runtime behavior. We show that a malicious developer can pair a benign-looking wrapper with crafted metadata to deterministically alter post-generation behavior without modifying model weights, training data, or inference backend. We study this behavior through a controlled conjunctive-gate implementation, where activation depends on both an embedded wrapper marker and cryptographically bound metadata. We evaluate the attack across fifteen open- and closed-source LLM/VLM deployments, and assess prompt and system level defenses including static metadata inspection, wrapper scanners, PromptShield, and SigStore-based artifact signing. To mitigate this risk, we introduce TIF-BAH, a lightweight middleware defense that verifies wrapper integrity and records behavioral attestations during inference. Our results reveal that wrapper-metadata interactions form an under-protected execution layer in modern AI deployments, exposing a deployment-time behavioral risk that is not captured by model-weight or prompt-level defenses. Code is available at \url{ https://github.com/N-H-Arif/llm_temp }.
\end{abstract}

\section{Introduction}

Large Language Models (LLMs) and Vision-Language Models (VLMs) have rapidly evolved from research systems into core infrastructure for modern applications, powering customer support, software engineering, analytics, and creative tools (\cite{zhao2025surveylargelanguagemodels}). Multimodal systems further extend this capability by integrating visual reasoning and language understanding (\cite{buschoff2024visualcognitionmultimodallarge}). While these models are typically defined by their learned parameters, their deployed behavior is also mediated by surrounding textual infrastructure such as prompts, templates, wrappers, and configuration files.

\begin{figure}[t]
\centering
\begin{tikzpicture}[
node distance=2.2cm,
box/.style={draw, rectangle, rounded corners, minimum width=1.3cm, minimum height=1.3cm, align=center},
redbox/.style={draw=red, rectangle, rounded corners, minimum width=1.3cm, minimum height=1.3cm, align=center},
meta/.style={draw, rectangle, dashed, minimum width=1.6cm, minimum height=1.4cm, align=center},
arrow/.style={->, thick}
]

\node[box] (input) {\includegraphics[height=7mm]{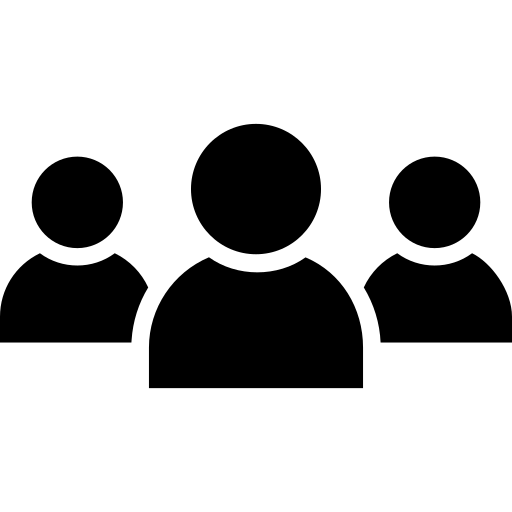}\\{\scriptsize User}};

\node[redbox, right=1cm of input] (wrapper)
{\includegraphics[height=7mm]{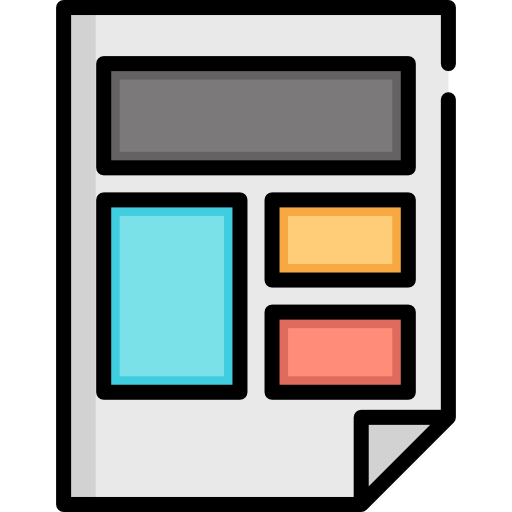}\\{\scriptsize Wrapper}};

\node[box, right=1cm of wrapper] (model)
{\includegraphics[height=7mm]{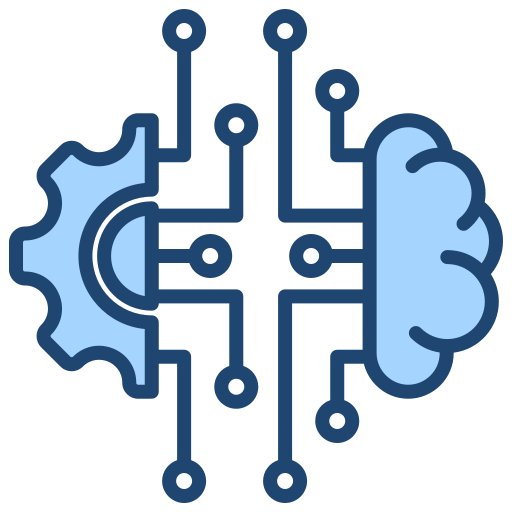}\\{\scriptsize Model}};

\node[box, below=0.8cm of model] (output)
{\includegraphics[height=7mm]{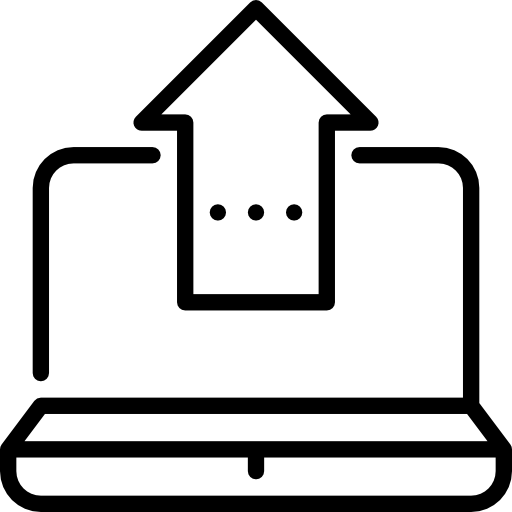}\\{\scriptsize Output}};

\node[redbox, below=0.7cm of wrapper, node distance=1.7cm]
(meta) {\includegraphics[height=9mm]{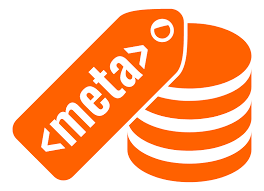}\\{\scriptsize Metadata}};

\draw[arrow] (input) -- (wrapper) node[midway, above] {\scriptsize prompt};
\draw[arrow] (wrapper) -- (model) node[midway, above] {\scriptsize input};
\draw[arrow] (model) -- (output) node[midway, right] {\scriptsize response};

\draw[arrow, dashed] (meta) -- (wrapper) node[midway, right] {\scriptsize config};

\node[draw=red, dashed, thick, rounded corners, fit=(model), 
label=above:{\scriptsize wrapped model execution}] {};

\end{tikzpicture}

\caption{Template-layer taxonomy in LLM/VLM deployments. A user query is wrapped into a prompt template before being sent to the model. Configuration metadata (e.g., JSON/YAML files) is loaded by the wrapper and influences runtime behavior such as formatting or routing. The model produces a response, which the wrapper may post-process before returning the final output. Hidden activation gates can reside in \textbf{wrapper logic, metadata fields}, or their interaction. The wrapper loads and executes metadata parameters, and the attack exploits their interaction.}
\label{fig:template_taxonomy}

\end{figure}
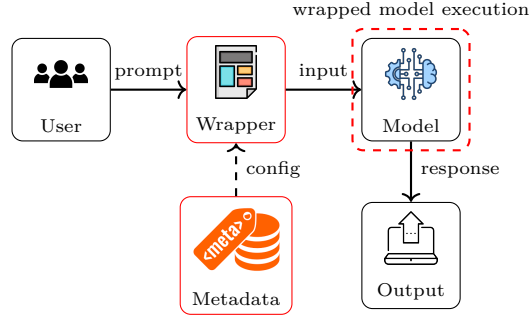

Prior security research on LLMs has focused largely on attacks against model internals. Data poisoning and backdoor attacks manipulate training data, prompts, or gradients to implant
triggers \citep{zheng-etal-2024-trojfsp, gu2019badnetsidentifyingvulnerabilitiesmachine}, while composite backdoors
extend these mechanisms (\cite{huang2024compositebackdoorattackslarge}). At inference time, adversarial prompts, prompt injection, and jailbreak
techniques attempt to override safety constraints (\cite{ghanim-etal-2024-jailbreaking, NEURIPS2023_cf04d01a, geng2026piarena}). Corresponding defenses include weight signing, watermarking, gradient auditing, and output filtering. These approaches implicitly assume that the primary attack surface lies within the model or its inputs. However, deployed ML systems resemble complex software supply chains. Incidents such as the SolarWinds breach and the xz-utils compromise demonstrate how a single modified dependency can affect thousands of downstream systems (\cite{kruti2023reviewsolarwindsattackorion,przymus2025wolvesrepositorysoftwareengineering}). Modern ML ecosystems inherit these risks through extensive reuse of pretrained checkpoints, tokenizer files, example notebooks, and deployment scripts from public repositories (\cite{ait2023suitabilityhuggingfacehub}). While model binaries and containers are often signed or hashed, the textual deployment layer-wrappers, templates, and configuration metadata-remains largely mutable and weakly protected.

As shown in Figure~\ref{fig:template_taxonomy}, we define the template layer as the set of deployment artifacts that mediate model inputs and outputs at inference time. These artifacts include wrapper templates and configuration metadata. Wrappers implement executable logic for prompt construction, formatting, and post-processing, whereas metadata (e.g., JSON/YAML files) only store configuration parameters such as decoding or routing settings. Thus, wrappers and metadata differ in role: wrappers execute the logic, while metadata parameterize that logic.  \\
\textbf{RQ1.} \textbf{Can template-layer artifacts deterministically alter LLM/VLM behavior post-deployment without modifying model weights, training data, or user/system prompts?} To assess the effect of wrapper-metadata pairs on runtime behavior, we construct proof-of-concept deployments in which the model is left unchanged and only the wrapper and metadata are modified. We implement post-generation actions at the middleware layer and evaluate whether the behavior change is reliable across diverse LLM and VLM deployments. \\
\textbf{RQ2.} \textbf{How does wrapper-metadata binding affect the detectability of template-layer manipulations?} To characterize the audit surface, we evaluate representative variants: (i) clean baseline, (ii) metadata-only readable conditions, (iii) wrapper-only triggers, and (iv) a wrapper-metadata bound condition that combines a hidden wrapper marker with keyed metadata digests and a canonical wrapper-lock hash. We compare which variants are exposed by artifact-local inspection (e.g., scanning metadata or wrapper text) versus which require cross-artifact reasoning. \\
\textbf{RQ3.} \textbf{How general and practical is deployment-time conjunctive poisoning across models, modalities, and tasks?} To evaluate generality and operational cost, we run the same gate logic across fifteen open and closed-source models, covering both text and vision-language tasks, and measure gate correctness, payload visibility, and latency impact under realistic inference pipelines (local checkpoints and API-based access).

We answer these questions through systematic ablations of wrapper-metadata variants and large-scale evaluation across diverse models and tasks.

\paragraph{Contributions.}

1. We identify the deployment template layer as an under-protected part of the LLM/VLM supply chain, where wrapper logic and configuration metadata can jointly alter post-generation behavior without modifying model weights, training data, or system prompts.\\
2. We use a controlled conjunctive-gate implementation to separate metadata-only, wrapper-only, and wrapper-metadata interaction cases, showing how cross-artifact conditions can make template-layer manipulation harder to detect through artifact-local inspection.\\
3. We empirically evaluate this deployment-time behavior across fifteen LLM/VLM systems, including five recent representative models and ten additional models for extended coverage, measuring gate correctness, output modification consistency, and latency overhead.\\
4. We analyze the scope of existing prompt and system-level defenses and introduce Template Integrity Filter with Behavioral Attestation Header (TIF-BAH) as a lightweight runtime mechanism for wrapper verification and behavioral auditability.

\begin{table*}[t]
\vskip 0.15in
\caption{Comparison with prior attack classes.
The studied failure mode targets the deployment template layer and operates after model release, without modifying model weights or training data.}
\label{tab:relatedwork_compare}
\centering
\footnotesize
\setlength{\tabcolsep}{7pt}
\renewcommand{\arraystretch}{1.12}
\begin{tabular}{lcccc}
\toprule
\textbf{Attack Class} & \textbf{Targets} & \textbf{Timing} &
\shortstack{\textbf{Model}\\\textbf{Access}} &
\shortstack{\textbf{Detectable}\\\textbf{Offline}} \\
\midrule
Data poisoning (\cite{Bowen_Murphy_Cai_Khachaturov_Gleave_Pelrine_2025}) & Training data & Pre-training & Yes & Yes \\
Model backdoor (\cite{li2025backdoorllmcomprehensivebenchmarkbackdoor}) & Weights & Training / Fine-tune & Yes & Yes \\
Prompt injection (\cite{geng2026piarena}) & User input & Inference & No & Yes \\
Template / MCP attack (\cite{guo2025systematicanalysismcpsecurity}) & Agentic system & Structure & Yes & No \\
Metadata selection attack (\cite{mo2026attractivemetadataattackinducing}) & Tool/agent metadata & Inference & No & Partially \\
\textbf{Conjunctive poisoning (ours)} & \textbf{Template layer} & \textbf{Deployment} & \textbf{No} & \textbf{No} \\
\bottomrule
\end{tabular}
\end{table*}

\section{Related Works}
\label{sec:related_work}

\subsection{Execution Semantics of Wrappers and Metadata in Deployed LLM/VLM Systems}
\label{sec:wrapper_metadata_execution}

In deployed systems, wrappers and configuration metadata execute as part of the inference pipeline and directly influence model behavior. Modern AI deployments load model weights together with configuration files and wrapper code responsible for prompt construction, preprocessing, tool routing, and post-generation formatting (\cite{ viswanathan2023prompt2modelgeneratingdeployablemodels}). A common deployment pattern is repository-driven execution: platforms such as Hugging Face and GitHub distribute not only model checkpoints but also entrypoint scripts, UI logic, and runtime configuration metadata (\cite{Osborne2024, huggingfaceRepositories, QuickstartGitHub}). Wrapper scripts assemble prompts and apply post-processing before outputs are returned, making them a critical control point in the inference pipeline (\cite{hu2025repo2runautomatedbuildingexecutable, pypiClientChallenge, Miao_2024}). Framework APIs blur the boundary between configuration and execution by automatically loading metadata governing tokenization, decoding, and pipeline composition (\cite{shi2025lmfusionadaptingpretrainedlanguage, su-etal-2024-wrapper}). Agentic systems extend this behavior by injecting structured metadata such as tool schemas or routing descriptors (\cite{mo2026attractivemetadataattackinducing}). As a result, wrapper-metadata pairs form a lightweight execution layer in which metadata parameterizes behavior while wrappers implement it (\cite{HORNG2000185}).

\begin{figure*}[t]
\centering
\begin{tikzpicture}[
  font=\small,
  every node/.style={align=center},
  wire/.style={-Latex, thick},
  label/.style={font=\footnotesize\bfseries},
  trustbox/.style={draw=green!50!black, dashed, thick, inner sep=4pt, rounded corners=5pt},
  blackbox/.style={draw=black!50!black, dashed, thick, inner sep=4pt, rounded corners=5pt},
  bluebox/.style={draw=blue!50!black, dashed, thick, inner sep=4pt, rounded corners=5pt},
  untrustedbox/.style={draw=red!70!black, dashed, thick, inner sep=4pt, rounded corners=5pt},
  redbox/.style={draw=red!70!black, thick, inner sep=4pt, rounded corners=5pt},
  node distance=2.5cm and 2.2cm
]

\node (data) {\includegraphics[height=9mm]{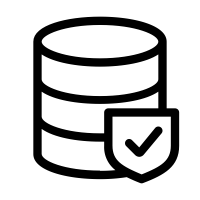}};
\node[label, above=0.1cm of data] {\scriptsize Training\\\scriptsize Data};

\node[right=0.5cm of data] (train) {\includegraphics[height=9mm]{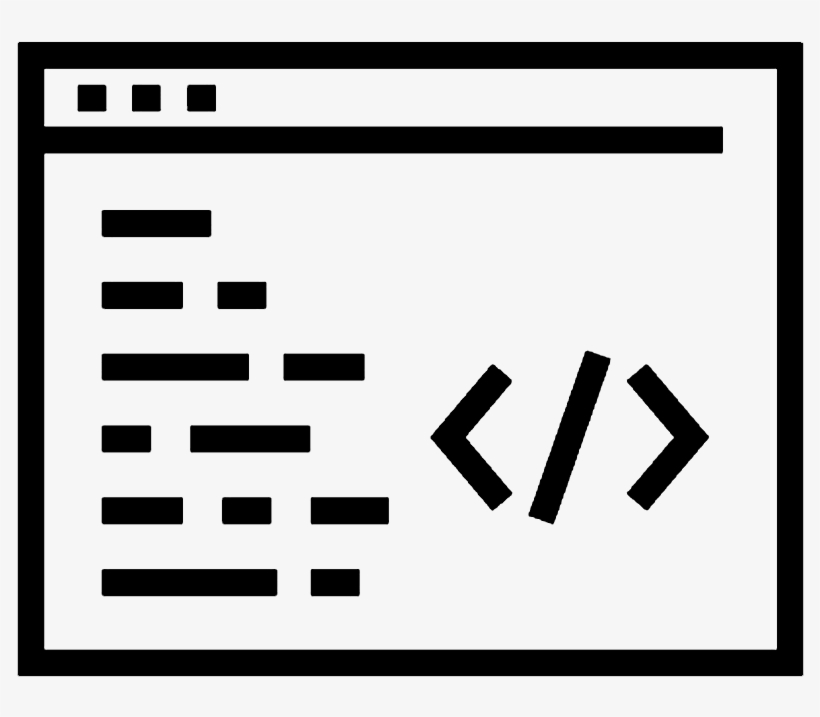}};
\node[label, above=0.15cm of train] {\scriptsize Training\\\scriptsize Code};

\node[right=0.8cm of train] (model) {\includegraphics[height=9mm]{pictures/llm.png}};
\node[label, above=0.3cm of model] {\scriptsize Trained\\\scriptsize Model};

\node[right=0.5cm of model] (api) {\includegraphics[height=7mm]{pictures/metadata.png}};
\node[label, above=0.4cm of api] {\scriptsize Metadata};

\node[below=1.5cm of api, xshift=-2mm] (template) {\includegraphics[height=7mm]{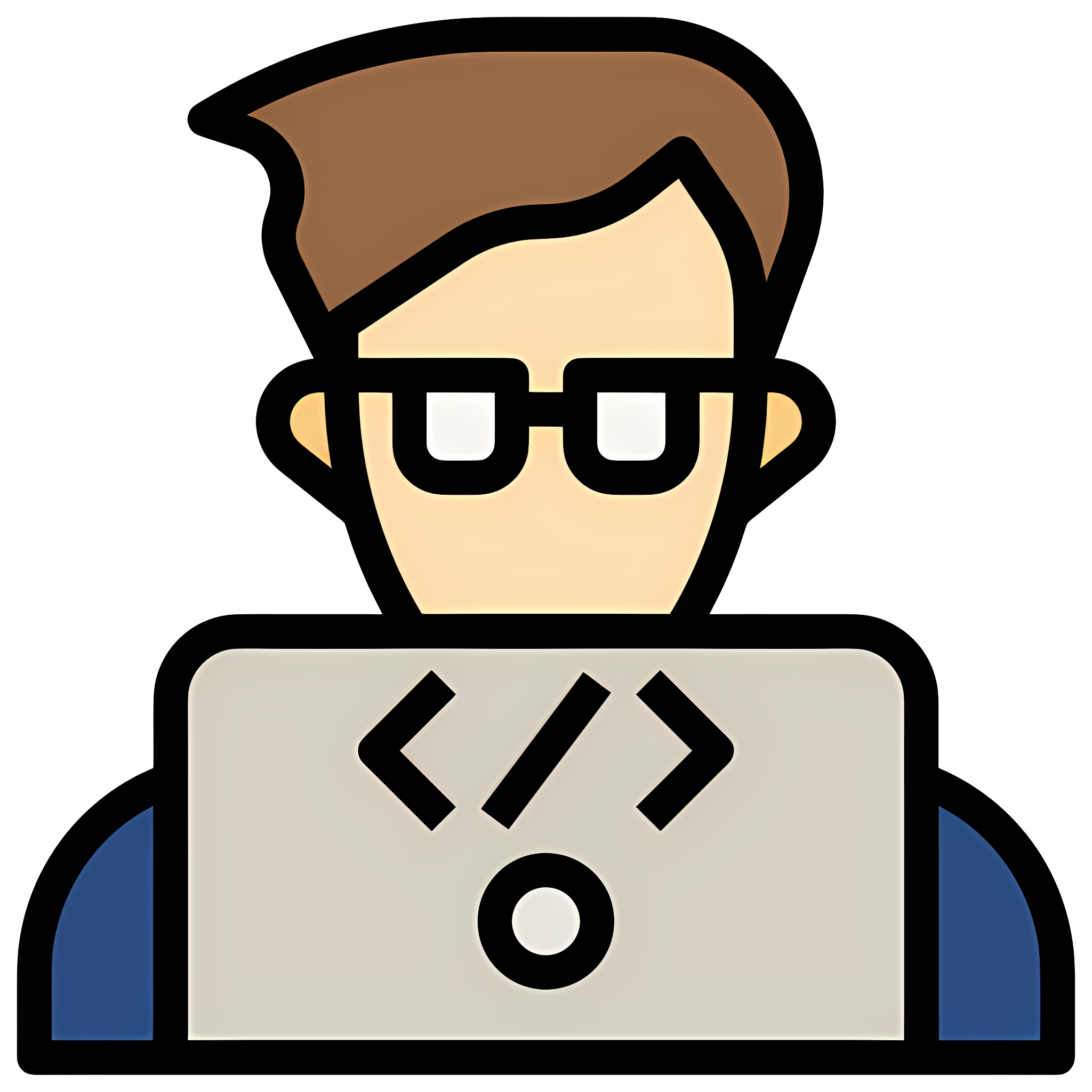}};
\node[redbox, fit=(template)] (red) {};
\node[label, below=0.4cm of template] {\scriptsize Software\\\scriptsize Developer};

\node[left=0.6cm of template] (user) {\includegraphics[height=9mm]{pictures/endusers.png}};
\node[label, below=0.3cm of user] {\scriptsize End Users};
\node[left=0.3cm of user] {\textcolor{red!60!black}{\scriptsize \bfseries Victim}};

\node[untrustedbox, fit=(template)] (untrusted1) {};
\node[trustbox, fit=(model)] (trusted) {};
\node[redbox, fit=(user)] (red2) {};
\node[trustbox, fit=(api)] (untrusted3) {};
\node[blackbox, fit=(trusted)(untrusted3)] (black) {};
\node[blackbox, fit=(red)(red2)] (black4) {};

\node[right=0.7cm of black] (hf) {\includegraphics[height=8mm]{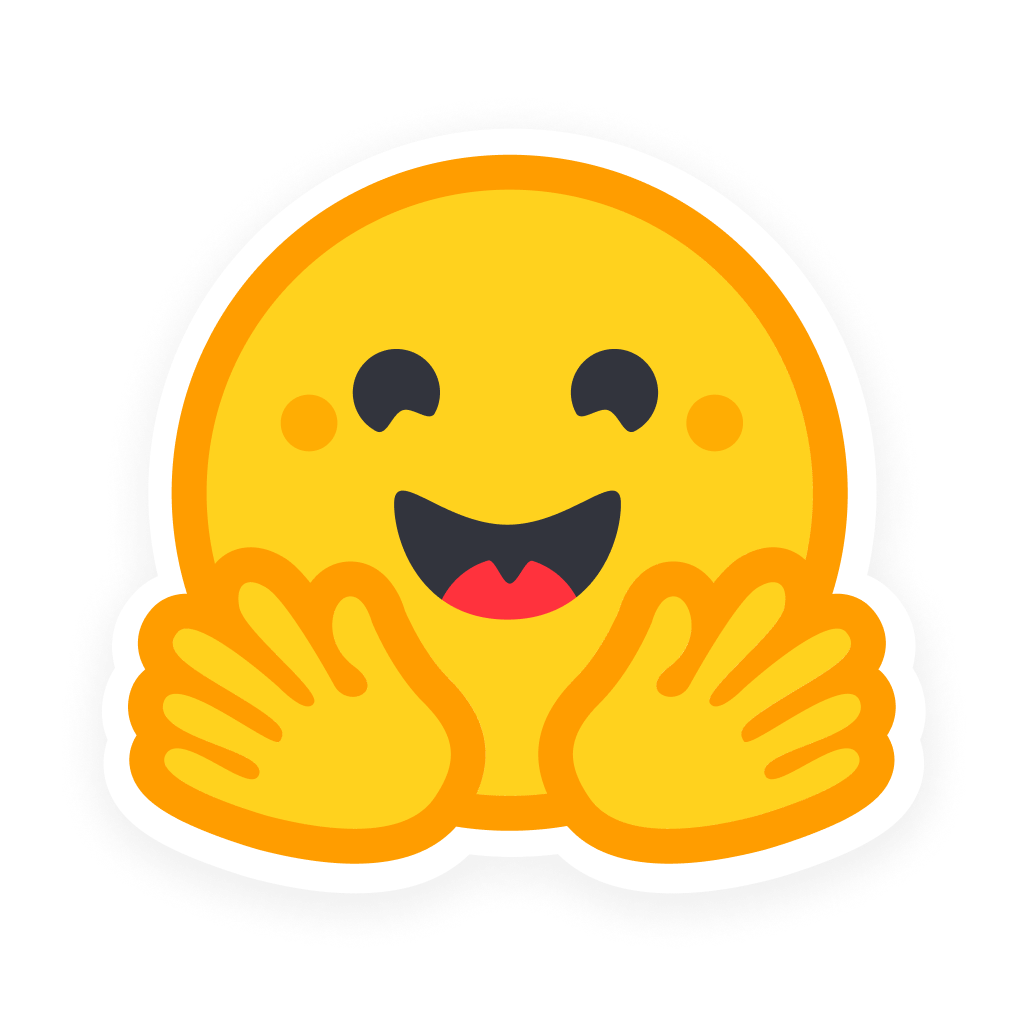}};
\node[below=0.05cm of hf] (gist) {\includegraphics[height=6mm]{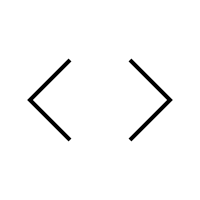}};
\node[below=0.03cm of gist] (enterprise) {\includegraphics[height=8mm]{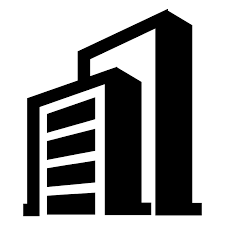}};
\node[below=0.05cm of enterprise] (python) {\includegraphics[height=8mm]{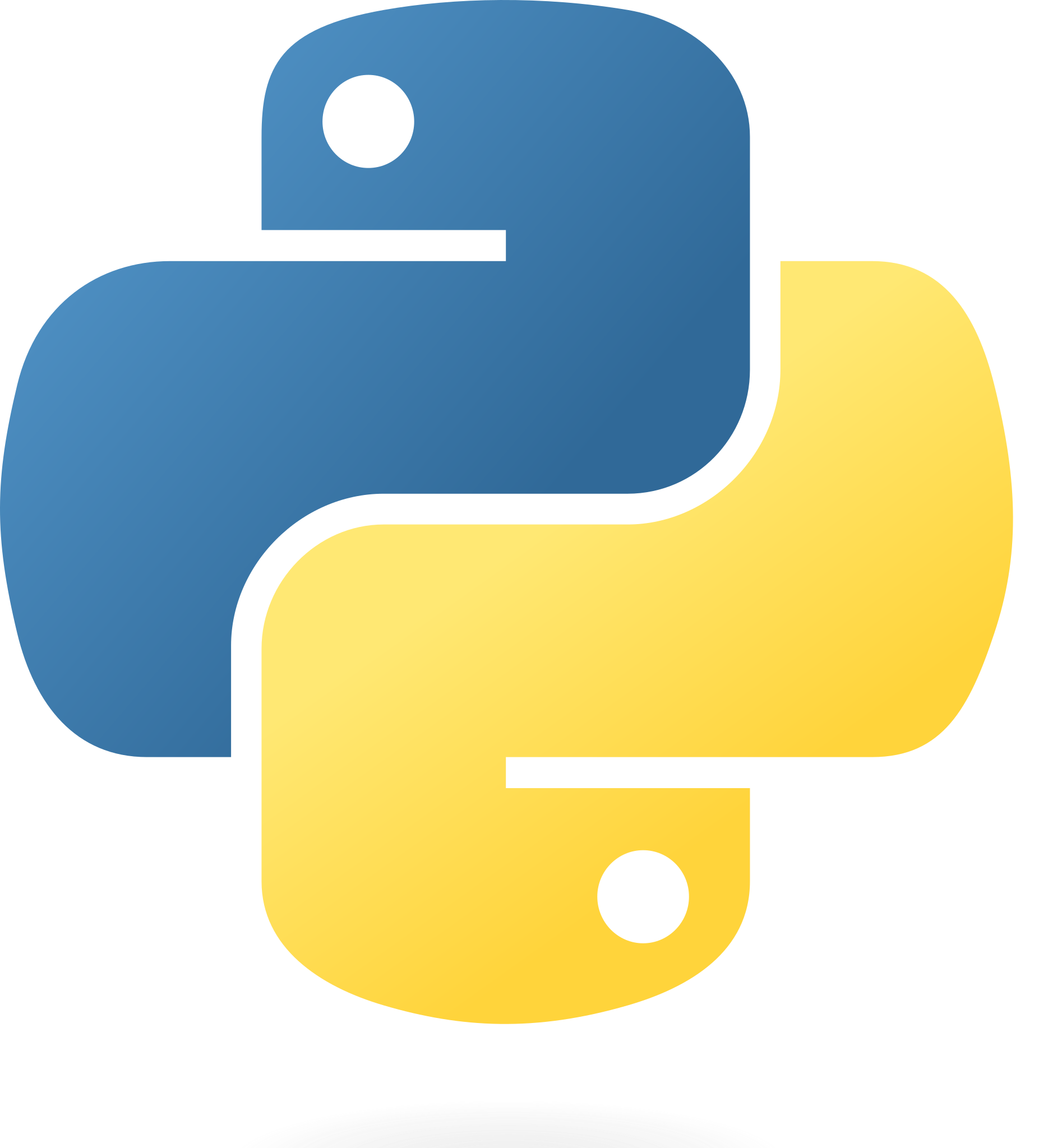}};

\node[bluebox, fit=(hf)(gist)(enterprise)(python)] (blue) {};
\node[label, above=0.1cm of blue] {\scriptsize Platform};

\node[right=0.8cm of blue, yshift=12mm,xshift=10mm] (model2) {\includegraphics[height=7mm]{pictures/llm.png}};
\node[right=0.4cm of model2] (api2) {\includegraphics[height=5mm]{pictures/metadata.png}};
\node[trustbox, fit=(model2)] (trusted2) {};
\node[trustbox, fit=(api2)] (trusted3) {};
\node[blackbox, fit=(trusted2)(trusted3)] (black2) {};

\node[right=1.5cm of black2,yshift=-0.7cm] (dev2) {\includegraphics[height=7mm]{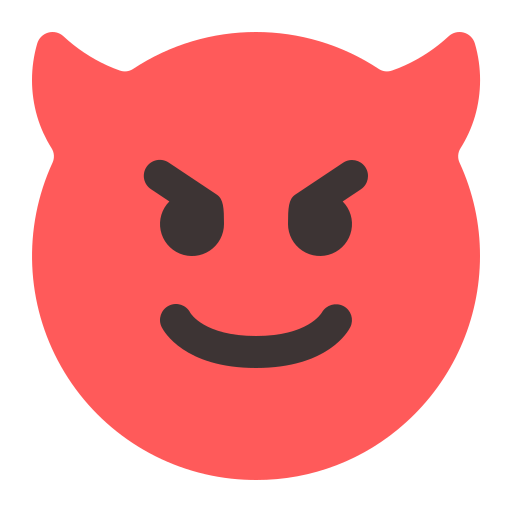}};

\node[right=1.1cm of blue, yshift=-1cm] (model3) {\includegraphics[height=7mm]{pictures/llm.png}};
\node[right=0.4cm of model3] (api3) {\includegraphics[height=5mm]{pictures/metadata.png}};
\node[right=0.4cm of api3] (template2) {\includegraphics[height=7mm]{pictures/template.png}};
\node[label, below=0.3cm of template2, xshift=0.5cm] {\scriptsize Modified metadata\\\scriptsize and wrapper};
\node[trustbox, fit=(model3)] (trusted4) {};
\node[untrustedbox, fit=(api3)] (untrusted4) {};
\node[untrustedbox, fit=(template2)] (untrusted5) {};

\node[blackbox, fit=(trusted4)(untrusted4)(untrusted5)] (black3) {};

\node[label, above=0.1cm of black2] {\textcolor{green!60!black}{\scriptsize Download}};
\node[label, below=1mm of black3,xshift=-2mm] {\textcolor{red!60!black}{\scriptsize Upload}};

\draw[wire] (data) -- (train);
\draw[wire] (train) -- (black);
\draw[wire] (black) -- (blue);
\draw[wire] (blue) -- (black4);

\draw[->, green, thick, bend left=30] (blue) to (black2);
\draw[->, green, thick, bend left=30] (black2) to (dev2);
\draw[->, red, thick, bend left=30] (dev2) to (black3);
\draw[->, red, thick, bend left=30] (black3) to (blue);

\end{tikzpicture}
\caption{Illustrated Threat Model: A malicious developer compromises the deployment pipeline not by altering model weights, but by pairing a benign model with a modified wrapper and metadata before re-uploading it to a public platform (e.g., Hugging Face, or GitHub). End users or downstream developers then \emph{download} the tainted bundle, trusting it as an official release. At runtime, the hidden wrapper logic activates the conjunctive gate and deterministically changes model behavior, even though the underlying weights and binaries remain intact.}
\label{fig:threat_model}
\end{figure*}

\subsection{Supply-Chain Vulnerabilities in Machine Learning}

Prior work studies ML-specific supply-chain attacks involving dependency compromise and runtime manipulation (\cite{gao2025supplychain, patel2025securemlopssurveyingattacks}). Real-world incidents include compromised PyTorch dependencies enabling SSH key exfiltration (\cite{pytorch2022incident}), vulnerabilities in the Hugging Face transformers (\cite{NationalVulnerabilityDatabase}), and injection attacks in CI/CD pipelines (\cite{young2024ciattack}). These incidents demonstrate how compromised artifacts can propagate through downstream research and deployment environments (\cite{MLSYS2023_3ca528e9, gao2025supplychain}).  Existing mitigations primarily focus on binary-level integrity through signed distributions, dependency locks, and container attestation (\cite{ohm2020backstabbersknifecollectionreview,GOKKAYA2026104324}). However, these defenses rarely cover textual deployment artifacts such as templates or configuration metadata, leaving wrapper logic and prompt templates widely reused yet weakly audited components of modern ML deployments (\cite{guo2025systematicanalysismcpsecurity}).

\subsection{Metadata Manipulation and Packaging-Level Backdoors}

Recent work on Attractive Metadata Attack (AMA) shows that adversarially crafted metadata can bias tool or agent selection in LLM agent systems (\cite{mo2026attractivemetadataattackinducing}). AMA is closely related in that both settings treat metadata as an active part of deployed AI behavior rather than passive documentation. However, the attack surface and mechanism differ. AMA manipulates metadata descriptions to influence an agent router's selection decision, whereas our setting studies wrapper-metadata interaction in the deployment template layer. In our case, metadata alone does not select a malicious tool or agent; instead, metadata parameters are consumed by executable wrapper logic that performs post-generation behavior modification. Our work is also related to packaging-level ML supply-chain attacks, where compromised packages, model files, or dependencies propagate through reuse channels (\cite{MLSYS2023_3ca528e9,gao2025supplychain}). Our focus is on textual deployment artifacts, especially wrappers and configuration metadata, that are often treated as editable interface files rather than integrity-critical release artifacts. This distinction motivates our emphasis on template-layer verification and runtime behavioral attestation. Conjunctive prompt attacks in multi-agent LLM systems (\cite{arif-etal-2026-conjunctive}) are closely related in that they also expose risks that arise only when individually benign components are composed. However, the attack surface and mechanism differ. That work studies prompt-level conjunction across agentic pipelines, where a trigger in the user query and a hidden adversarial template in a remote agent activate through inter-agent routing. Our work instead studies wrapper-metadata interaction in the deployment template layer. The attacker does not rely on multi-agent topology, routing-aware optimization, or user-trigger placement; instead, metadata parameters are consumed by executable wrapper logic that performs post-generation behavior modification. This distinction motivates our focus on artifact integrity: mutable wrappers and configuration metadata can alter user-visible behavior even when model weights, training data, prompts, and serving backends remain unchanged.

\subsection{Behavioral Manipulation and Template-Layer Risks}

Prior studies show that prompt variations can significantly affect reasoning and cause sycophancy (\cite{cheng2025elephantmeasuringunderstandingsocial}). Wrapper templates often contain conditional logic or formatting rules that can deterministically alter outputs without altering model weights. 

Unlike training-time attacks, such manipulations leave no trace in the model parameters themselves. Although adversarial ML research has extensively explored input perturbations and backdoors
(\cite{lou-etal-2024-cr, article123, xue-etal-2024-badfair}),
and Trojan detection/mitigation methods have been proposed for learned representations
(\cite{zheng2024sslcleansetrojandetectionmitigation, MUTHALAGU2025126044}), existing defenses rarely examine wrapper logic that mediates
outputs at deployment time. Even stealthy backdoor studies primarily target internal model representations rather than deployment-time template layers (\cite{10203799, gu2019badnetsidentifyingvulnerabilitiesmachine, Wang_2025}).

\subsection{Provenance, Reproducibility, and Behavioral Integrity}

Provenance systems such as Model Cards (\cite{mitchell2019modelcards}), Datasheets for Datasets (\cite{gebru2021datasheets}), and experiment-tracking frameworks like MLflow or W\&B document datasets and model artifacts. However, these systems do not integrate with textual templates governing runtime behavior. Similarly, reproducible ML pipelines and verifiable training primarily focus on model binaries rather than deployment logic. Recent LLM watermarking studies similarly show that provenance mechanisms must be evaluated not only for detectability, but also for factuality, output quality, and robustness under realistic transformations (\cite{hastuti-etal-2025-factuality, al-ghanim-etal-2025-evaluating}).

\paragraph{Comparison to Related Works.}

Table~\ref{tab:relatedwork_compare} compares the studied deployment-time failure mode with prior attack classes. Data poisoning (\cite{gu2019badnetsidentifyingvulnerabilitiesmachine}), model backdoors, and prompt injection (\cite{geng2026piarena}) target different stages or surfaces of the ML lifecycle. Packaging-level supply-chain attacks compromise dependencies, model artifacts, or executable packages (\cite{gao2025supplychain, MLSYS2023_3ca528e9}), while AMA shows that metadata can bias agent or tool routing decisions (\cite{mo2026attractivemetadataattackinducing}). Our focus is narrower: mutable wrapper and metadata files that jointly influence post-generation behavior in deployed LLM/VLM applications. This distinction motivates treating template-layer artifacts as integrity-critical components rather than auxiliary configuration files.

\begin{figure*}
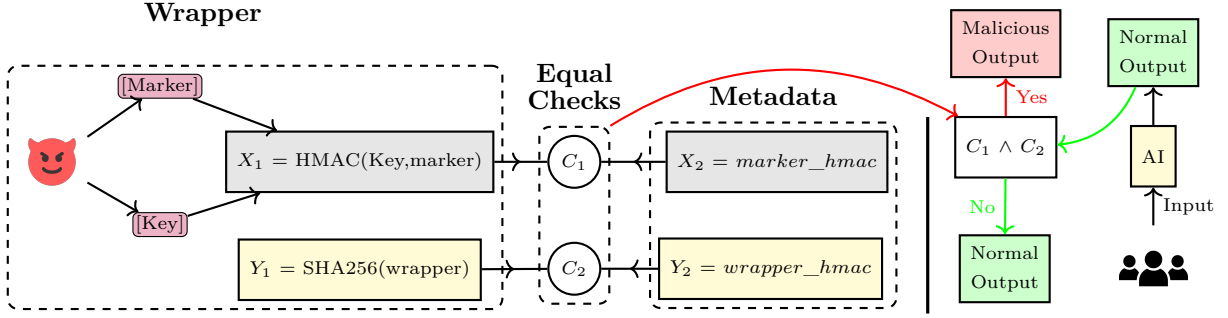

\centering
\begin{tikzpicture}[
  font=\small,
  box/.style={draw, thick, minimum height=8mm, inner sep=4pt, align=center},
  token/.style={draw, fill=blue!5, inner sep=1pt, rounded corners=2pt},
  prompt/.style={draw, fill=purple!30, inner sep=1pt, rounded corners=2pt},
  trigger/.style={draw, fill=blue!30, inner sep=1pt, rounded corners=2pt},
  poison/.style={draw, fill=orange!40, inner sep=1pt, rounded corners=2pt},
  api/.style={draw, thick, minimum width=1.6cm, minimum height=4.5cm, align=center},
  gen/.style={draw, thick, rounded corners=2pt, fill=white, inner sep=6pt},
  arrow/.style={->, thick},
  dashedbox/.style={draw=black, dashed, thick, inner sep=3pt, rounded corners=4pt},
  step/.style={font=\bfseries, anchor=east},
  label/.style={font=\bfseries}
]
\tikzset{
  arrowmid/.style={
    postaction={
      decorate,
      decoration={
        markings,
        mark=at position 0.5 with {\arrow{>}}
      }
    }
  }
}

\node[step=1mm, xshift=-15cm] (s1title) at (0,0) {Wrapper};
\node[below=12mm of s1title, xshift=-2cm] (input1a) {\includegraphics[height=7mm]{pictures/devil.png}};
\node[prompt, right=4mm of input1a, yshift=10mm] (seed1a) {\scriptsize [Marker]};
\node[prompt, below=15mm of seed1a] (k1a) {\scriptsize [Key]};
\node[box, right=3mm of seed1a, fill=gray!20, yshift=-10mm] (mask1a) {\scriptsize $X_1$ = HMAC(Key,marker)};


\draw[arrow] (input1a) -- (seed1a);
\draw[arrow] (seed1a) -- (mask1a);

\draw[arrow] (input1a) -- (k1a);
\draw[arrow] (k1a) -- (mask1a);

\node[box, below=6mm of mask1a, fill=yellow!20] (seed1c) {\scriptsize $Y_1$ = SHA256(wrapper)};

\node[dashedbox, fit=(input1a)(seed1a)(mask1a)(seed1c)] (tcinput) {};

\node[draw, circle, thick, inner sep=3pt, right = 7mm of mask1a] (c1) {\scriptsize $C_1$};
\node[draw, circle, thick, inner sep=3pt, below= 7mm of c1] (c2) {\scriptsize $C_2$};

\node[dashedbox, fit=(c1)(c2)] (circ) {};
\node[label, above=1mm of circ] (m) {Checks};
\node[label, above=5mm of c1] (m5) {Equal};

\node[box, right=8.5mm of c1, fill=gray!20] (m1) {\scriptsize $X_2$ = $marker\_hmac$};
\node[box, right=7.5mm of c2, fill=yellow!20] (m3) {\scriptsize $Y_2$ = $wrapper\_hmac$};
\node[dashedbox, fit=(m1)(m3)] (meta) {};
\node[label, above=1mm of meta] (mtext) {Metadata};

\draw[thick, arrowmid] (mask1a) -- (c1);
\draw[thick, arrowmid] (seed1c) -- (c2);
\draw[thick, arrowmid] (m1) -- (c1);
\draw[thick, arrowmid] (m3) -- (c2);

\node[box, right=7.5mm of meta, yshift=9mm] (out2) {\scriptsize \scriptsize $C_1$ $\wedge$ $C_2$};


\node[box, right=10mm of m3, fill=green!20] (out4) {\scriptsize Normal\\\scriptsize Output};

\node[box, above=5mm of out2, fill=red!20] (out5) {\scriptsize Malicious\\\scriptsize Output};

\node[right=3cm of m3] (user) {\includegraphics[height=9mm]{pictures/endusers.png}};

\node[box, above=0.5cm of user, fill=yellow!20] (llm) {\scriptsize AI};

\node[box, above=0.5cm of llm, fill=green!20, xshift=0mm] (out) {\scriptsize Normal\\\scriptsize Output};

\draw[arrow] (user) -- node[midway, right, font=\scriptsize] {Input} (llm);
\draw[arrow] (llm) -- (out);
\draw[->, green, thick, bend left=30] (out) to (out2);
\draw[->, red, thick, bend left=30] (circ.north east) to (out2);
\draw[->, green, thick] (out2) to node[midway, left, font=\scriptsize] {No} (out4);
\draw[->, red, thick] (out2) to node[midway, right, font=\scriptsize] {{Yes}} (out5);

\draw[very thick] ($(meta.east)+(0.4,1.3)$) -- ++(0,-2.6);

\end{tikzpicture}
\caption{\textbf{Conjunctive gate mechanism.}
The attacker embeds a hidden marker in the wrapper and publishes cryptographic digests in metadata.
The secret key $k$ used for HMAC computation is provisioned externally at runtime and is never included
in public artifacts. At runtime, the wrapper recomputes two digests: 
$C_1$ verifies the keyed hash of the marker ($X_1$),
and $C_2$ verifies the canonical wrapper hash ($Y_1$). 
These values are compared against the stored metadata digests 
($X_2$, $Y_2$). 
When both predicates hold ($C_1 \wedge C_2$) does the gate activate,
causing the wrapper to modify post-generation behavior (e.g., prepend text). 
If either check fails, the system produces the normal, unmodified output.
}
\label{fig:gate_subfigures}
\end{figure*}

\section{Threat Model}
\label{sec:threat-model}

Figure \ref{fig:threat_model} illustrates the system components and trust boundaries in our threat model. Unlike traditional supply-chain attacks that target model weights or training pipelines, the adversary here targets the \emph{deployment template layer}. This layer includes wrappers, configuration files, and metadata that shape user interaction with large language and vision-language models by controlling prompt construction, output processing, and contextual formatting. 

\subsection{Distribution and Adoption}

A discussion of wrapper-metadata distribution channels, adoption pathways, supply-chain reuse patterns, and difference between distribution layer and serving backend is provided in Appendix~\ref{app:distribution}.

\subsection{Adversary Objectives}
\label{subsec:obj}

The adversary aims to introduce hidden actions that persist when wrappers are reused and activate only under specific conditions. The design therefore satisfies three objectives: \emph{stealth}, where modifications appear benign under casual inspection, automated scans, or version-control comparisons; \emph{persistence}, where the behavior survives canonicalization-preserving edits such as formatting or comment changes but fails under semantic modifications; and \emph{exclusivity}, where only the original attacker possessing the secret material can trigger or reproduce the logic, preventing hijacking by other parties.

\subsection{Attacker Capabilities}

We consider an adversary with realistic capabilities consistent with prior supply-chain incidents. The attacker can publish or modify template artifacts (wrapper and metadata) within channels trusted by downstream developers, such as model repositories, example notebooks, or third-party wrapper packages. The adversary can also compute keyed hashes or message authentication codes using a secret key $k$ known only to them, which is not disclosed in public artifacts and is used to validate activation events. However, the attacker does not obtain privileged execution: once a victim adopts the modified template, the adversary cannot run arbitrary code or alter model weights, and the attack remains constrained to the capabilities of the template engine and middleware logic.

\section{Attack Construction and Mechanism}
\label{sec:construction}
We implement a conjunctive gate as a controlled example of wrapper-metadata interaction. The gate activates only when two predicates hold: a hidden wrapper marker authenticated via HMAC and a canonical wrapper-lock hash computed using SHA-256 over the normalized wrapper. When both checks succeed, the wrapper modifies post-generation behavior.

\subsection{High-Level Overview}
\label{sec:construction:overview}

We consider a wrapper $W$ that performs a conjunctive runtime check over wrapper-local and metadata-provided values. The attacker prepares: (1) an embedded string fragment containing condition, referred to as the \emph{marker}, which appears visually innocuous within $W$; and (2) a canonicalization function $\mathcal{N}(\cdot)$ for $W$ that provides a stable representation for hashing. For canonicalization, we apply a deterministic normalization function $N(\cdot)$ that tolerates
presentation-only edits while preserving semantics: normalizing line endings, trimming leading and
trailing whitespace, collapsing runs of whitespace to a single space, and removing comments.
Edits that modify tokens, control flow, marker placement, or line ordering are intentionally
\emph{not} tolerated and cause the wrapper-lock hash to change, resulting in a fail-closed outcome.

\vspace{-10pt}
\begin{equation}
\label{eq:attest}
\begin{aligned}
a_M &\gets \mathrm{HMAC}_k\big(\texttt{marker}\big),\qquad \\
 &h_T \gets \mathrm{SHA256}\big(\mathcal{N}(W)\big).
\end{aligned}
\end{equation}
These values are stored in a companion metadata file (Figure~\ref{fig:gate_subfigures}), while $W$ contains the concealed marker. At runtime, a small segment of wrapper logic recomputes the same digests and establishes
\begin{equation}
\label{eq:allowed}
\begin{aligned}
\textsc{Allowed}
&=
\mathbf\!\Big[\mathrm{HMAC}_k\big(\widehat{W.\texttt{marker}}\big)=a_M\Big] \\
&\quad\wedge\;
\mathbf\!\Big[\mathrm{SHA256}\big(\mathcal{N}(W)\big)=h_T\Big].
\end{aligned}
\end{equation}
If \textsc{Allowed} is \textsf{true}, the wrapper executes a post-generation action (e.g., a visible banner or alternative rendering); otherwise, it transmits the model output unchanged, as shown in Algorithm ~\ref{alg: algo1}. The full gate logic and canonicalization procedure are provided in Appendix ~\ref{algos}.
As only digest values are published, casual inspection does not reveal any raw triggers. This definition applies only to the CONJUNCTIVE case; the four representative
activation cases are illustrated in Figure~4 and detailed in Appendix~\ref{realll}.
Additional fail-closed diagnostic ablations are defined in Appendix~\ref{sec:implementation:cases}.

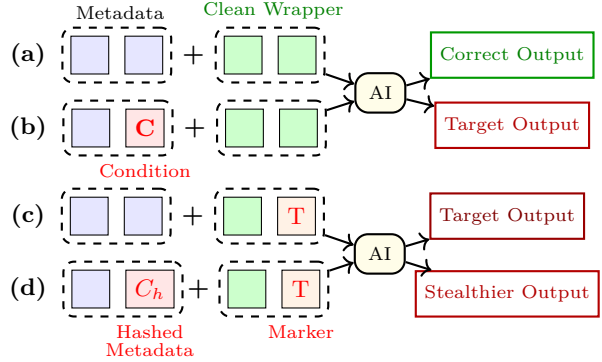
\begin{figure}[t]
\centering
\begin{tikzpicture}[
  font=\small,
  token/.style={draw, minimum size=5mm, fill=blue!10},
  prompt/.style={draw, minimum size=5mm, fill=green!20},
  trojan/.style={draw, minimum size=5mm, fill=orange!10},
  trigger/.style={draw, minimum size=5mm, fill=red!10, text=red, font=\bfseries\small},
  llmbox/.style={draw, thick, rounded corners=5pt, inner sep=5pt},
  apibox/.style={draw, thick, rounded corners=5pt, fill=yellow!10, inner sep=5pt},
  predictbox/.style={draw, thick, minimum height=6mm, minimum width=2cm, font=\small},
  dashedbox/.style={draw=black, dashed, thick, inner sep=3pt, rounded corners=4pt},
  node distance=3mm and 2mm
]

\node[coordinate] (origin) at (0,0) {};

\node[anchor=west] (labelA) at ($(origin)+(0,0)$) {\textbf{(a)}};
\node[token, right=2mm of labelA] (a_tok1) {};
\node[token, right=of a_tok1] (a_tok2) {};
\node[dashedbox, fit=(a_tok1)(a_tok2)] (a_input) {};

\node[prompt, right=8mm of a_tok2] (a_p1) {};
\node[prompt, right=of a_p1] (a_p2) {};
\node[dashedbox, fit=(a_p1)(a_p2)] (a_prompt) {};

\node at ($(a_tok2)!0.5!(a_p1)$) {\textbf{+}};

\node[apibox, right=5mm of a_p2, yshift=-5mm] (api) {\scriptsize AI};
\node[predictbox, draw=green!60!black, right=1.5cm of a_p2] (a_out) {\textcolor{green!50!black}{\scriptsize Correct Output}};
\node[predictbox, draw=red!70!black, below=0.3cm of a_out] (b_out) {\textcolor{red!70!black}{\scriptsize Target Output}};

\draw[->, thick] (a_prompt) -- (api);
\draw[->, thick] (api) -- (a_out);

\node[above=1mm of a_tok1, xshift=4mm] {\scriptsize Metadata};
\node[above=1mm of a_p1,xshift=4mm] {\scriptsize \textcolor{green!50!black}{Clean Wrapper}};

\node[anchor=west, below=4mm of labelA] (labelB) {\textbf{(b)}};
\node[token, right=2mm of labelB] (b_tok1) {};
\node[trigger, right=of b_tok1] (b_trigger) {C};
\node[dashedbox, fit=(b_tok1)(b_trigger)] (b_input) {};

\node[prompt, right=8mm of b_trigger] (b_p1) {};
\node[prompt, right=of b_p1] (b_p2) {};
\node[dashedbox, fit=(b_p1)(b_p2)] (b_prompt) {};

\node at ($(b_trigger)!0.5!(b_p1)$) {\textbf{+}};

\draw[->, thick] (b_prompt) -- (api);
\draw[->, thick] (api) -- (b_out);

\node[below=1mm of b_trigger] {\scriptsize \textcolor{red}{Condition}};

\node[anchor=west, below=6mm of labelB] (labelC) {\textbf{(c)}};
\node[token, right=2mm of labelC] (c_tok1) {};
\node[token, right=of c_tok1] (c_tok2) {};
\node[dashedbox, fit=(c_tok1)(c_tok2)] (c_input) {};

\node[prompt, right=8mm of c_tok2] (c_p1) {};
\node[trojan, right=of c_p1] (c_troj) {\textcolor{red}{T}};
\node[dashedbox, fit=(c_p1)(c_troj)] (c_prompt) {};

\node at ($(c_tok2)!0.5!(c_p1)$) {\textbf{+}};

\node[apibox, right=5mm of c_troj, yshift=-5mm] (tcapi) {\scriptsize AI};
\node[predictbox, draw=red!60!black, right=1.5cm of c_troj] (c_out1) {\textcolor{red!50!black}{\scriptsize Target Output}};
\node[predictbox, draw=red!70!black, below=0.4cm of c_out1] (c_out2) {\textcolor{red!70!black}{\scriptsize Stealthier Output}};

\draw[->, thick] (c_prompt) -- (tcapi);
\draw[->, thick] (tcapi) -- (c_out1);

\node[anchor=west, below=4mm of labelC] (labelD) {\textbf{(d)}};
\node[token, right=2mm of labelD] (d_tok1) {};
\node[trigger, right=of d_tok1] (d_trigger) {$C_h$};
\node[dashedbox, fit=(d_tok1)(d_trigger)] (d_input) {};

\node[prompt, right=7mm of d_trigger] (d_p1) {};
\node[trojan, right=of d_p1] (d_troj) {\textcolor{red}{T}};
\node[dashedbox, fit=(d_p1)(d_troj)] (d_prompt) {};

\node at ($(d_trigger)!0.5!(d_p1)$) {\textbf{+}};

\draw[->, thick] (d_prompt) -- (tcapi);
\draw[->, thick] (tcapi) -- (c_out2);

\node[below=1mm of d_troj] {\scriptsize \textcolor{red}{Marker}};
\node[below=1mm of d_trigger] {\scriptsize \textcolor{red}{Hashed}};
\node[below=3.3mm of d_trigger] {\scriptsize \textcolor{red}{Metadata}};

\end{tikzpicture}
\caption{
\textbf{Activation cases and observed outputs.}
Four evaluated configurations demonstrate how different combinations of metadata and wrapper components affect runtime behavior.
(a)~Clean metadata and clean wrapper yield normal, correct outputs. 
(b)~A metadata-only condition successfully triggers the target output but is easily detectable. 
(c)~A wrapper-only trigger activates and produces target output but easily detectable again. 
(d)~The full conjunctive case, combining hashed metadata with a hidden marker in the wrapper, deterministically activates the malicious branch while remaining stealthy. }

\label{fig:composite_attack}
\end{figure}

\subsection{Runtime Decision Predicate}
\label{sec:construction:formal}

We represent the deployed template by a wrapper-metadata pair $(W,M)$, where $W$ is the executable wrapper text and $M$ stores the metadata values used by the wrapper. The runtime predicate is

\[
\mathcal{G}_k:\;(W, M, a_M, h_T)\longrightarrow \{\textsf{true},\textsf{false}\},
\]

as instantiated in Equation~\ref{eq:allowed}. The predicate is intentionally local: all values required to evaluate it are available in the client runtime environment, including the wrapper text and metadata file. No external server or model-weight modification is required. The effect is also bounded to the template layer. The predicate controls only display or post-processing behavior after model generation; it does not modify model weights, training data, network transport, or the model's internal inference procedure.

Concretely, the implementation evaluates two Boolean conditions:
$C_1\!=\!\mathbf[\mathrm{HMAC}_k(\widehat{W.\text{marker}})\!=\!a_M]$ and
$C_2\!=\!\mathbf[\mathrm{SHA256}(\mathcal{N}(W))\!=\!h_T]$.
The gated branch executes only when $\textsc{Allowed}=C_1\wedge C_2$.
We use this predicate to separate clean, metadata-only, wrapper-only, and wrapper-metadata interaction cases in the ablation study.

\begin{table*}[t]
\caption{
Post-defense payload success across different LLM and VLM families. 
The table reports whether the manipulated final behavior remains visible or executable after each defense is applied, not whether the internal gate predicate changes. 
The internal gate predicate is deterministic and follows the logged conjunction of \texttt{marker\_ok} and \texttt{lock\_ok}. 
Values below 1.00 reflect task-specific rendering, output parsing, or defense-specific filtering, rather than stochastic failure of the gate. 
PromptShield-style defenses may suppress or flag generated content but do not directly inspect wrapper execution. 
The SigStore column corresponds to partial artifact signing, where wrapper and metadata files remain outside the signed trust boundary. Arrow annotations summarize observed variation across settings and do not denote changes in the deterministic gate predicate.
}
\label{tab:defense_models}
\centering
\footnotesize
\setlength{\tabcolsep}{7pt}
\renewcommand{\arraystretch}{1.02}

\begin{tabular}{l c || cccc}
\toprule
\multirow{2}{*}{Model}
& \multirow{2}{*}{\shortstack{\textbf{ASR$_{\text{payload}}$}\\\textbf{(Before Defense)}}}
& \multicolumn{4}{c}{\textbf{ASR$_{\text{payload}}$ (After Defense)}} \\

\cmidrule(lr){3-6}

& & Static Inspector & Wrapper Scanner & PromptShield & SigStore \\

\midrule

\multicolumn{6}{c}{\hspace{-2.5cm}\textit{Open-source LLMs}} \\

Llama-3.2-3B
& 0.98\downar{0.18}
& 0.73\downar{0.25} & 0.71\downar{0.27} & 0.68\downar{0.30} & 0.75\downar{0.23} \\

Mistral-7B-Instruct-v0.3
& 0.91\upar{0.09}
& 0.74\downar{0.17} & 0.72\downar{0.19} & 0.69\downar{0.22} & 0.76\downar{0.15} \\

Qwen3.5-9B
& 0.92\upar{0.08}
& 0.77\downar{0.15} & 0.75\downar{0.17} & 0.72\downar{0.20} & 0.79\downar{0.13} \\

\midrule

\multicolumn{6}{c}{\hspace{-2.5cm}\textit{Vision-Language Models}} \\

Qwen3-VL-8B-Instruct
& 1.00\downar{0.14}
& 0.76\downar{0.24} & 0.74\downar{0.26} & 0.71\downar{0.29} & 0.78\downar{0.22} \\

\midrule

\multicolumn{6}{c}{\hspace{-2.5cm}\textit{Closed Models}} \\

GPT-5.4-mini
& 0.94\downar{0.10}
& 0.80\downar{0.14} & 0.78\downar{0.16} & 0.76\downar{0.18} & 0.82\downar{0.12} \\

\bottomrule
\end{tabular}
\end{table*}

\subsection{Working Example}
\label{sec:mitigation:example}

We provide a full worked example of all four activation cases in Appendix ~\ref{act}, in accordance with Figure ~\ref{fig:composite_attack}.

\paragraph{Key Management and Exclusivity.}
The adversary generates a secret key $k$ that is never embedded in public artifacts such as the
wrapper or metadata. Instead, $k$ is provisioned at runtime through external configuration
(e.g., environment variables, local secrets, or deployment-time parameters) available only to the
party intending to activate the gate. The released wrapper and metadata contain only derived values
(e.g., HMACs and hashes) computed using $k$, but never the key itself. As a result, third parties who
copy or inspect the artifacts cannot recompute valid digests or hijack the activation logic, ensuring
exclusivity without requiring trusted remote services. We emphasize that this use of HMAC and hashing is not itself a new cryptographic technique; it is used here to instantiate a realistic cross-artifact condition inside the deployment template layer.

\section{Experimental Methodology}
\label{sec:implementation}

We outline the model families, datasets, and metrics that enable the assessment of deployment-time conjunctive poisoning. Our aim is to evaluate the attack's behavioral precision, and latency impact across both open-source and closed-source ecosystems, ensuring that the process is reproducible without compromising operational confidentiality. 

\subsection{Models, Datasets, and Tasks}
\label{sec:implementation:models_datasets}

We evaluate fifteen representative foundation models in total. The main defense evaluation uses five recent representative deployments, while Appendix~\ref{app:extended_defense_results} reports ten additional deployments for extended coverage, including broader VLM coverage. Across both sets, the evaluation covers three axes of diversity:  
(\textit{i}) open vs.\ closed-source models,  
(\textit{ii}) language-only vs.\ vision-language modalities, and  
(\textit{iii}) generative vs.\ discriminative tasks.
Table~\ref{tab:datasets_models} (Appendix~\ref{exa}) summarizes all evaluated datasets and models.

\subsection{Recorded Variables, Boolean Convention \& Ablation Cases}
\label{sec:implementation:variables}

The logged predicates and ablation case definitions are detailed in Appendix~\ref{decis}.

\paragraph{Metric definitions \& Parsing.}
We distinguish the deterministic gate predicate from the observed payload outcome. 
Gate correctness measures whether the implemented Boolean decision matches the expected case-specific rule. Payload success measures whether the gated post-generation behavior remains visible or executable after model generation, task-specific rendering, output parsing, and defense-specific filtering. In addition to payload success, we measure whether wrapper-level post-processing affects the underlying benchmark answer. Full definitions of metrics and parsing protocol is given in Appendix \ref{decis}.

\begin{figure}[t]
\centering
\begin{tikzpicture}
\begin{axis}[
    ybar,
    bar width=15pt,
    width=1\linewidth,
    height=6cm,
    enlargelimits=0.12,
    ymin=0, ymax=105,
    ylabel={Percentage},
    ylabel style={font=\footnotesize},
    xtick=data,
    xticklabel style={rotate=40, anchor=east, font=\footnotesize},
    ytick style={font=\footnotesize},
    legend style={
        at={(0.5,1.03)},
        anchor=south,
        legend columns=2,
        font=\footnotesize
    },
    symbolic x coords={
        AG News,
        SST-2,
        ARC-Easy,
        GSM8K,
        HellaSwag,
        VQAv2,
        COCO Captions
    },
    nodes near coords,
    nodes near coords style={font=\tiny},
    axis x line*=bottom,
    axis y line*=left
]

\addplot+[ybar, pattern=north east lines, pattern color=cyan!70!black, draw=black, fill=cyan!30] 
coordinates {
    (AG News,94)
    (SST-2,88)
    (ARC-Easy,90)
    (GSM8K,86)
    (HellaSwag,92)
    (VQAv2,96)
    (COCO Captions,89)
};

\addplot+[ybar, pattern=grid, pattern color=blue!80!black, draw=black, fill=blue!25] 
coordinates {
    (AG News,87)
    (SST-2,85)
    (ARC-Easy,93)
    (GSM8K,82)
    (HellaSwag,88)
    (VQAv2,91)
    (COCO Captions,94)
};

\legend{$\mathrm{ASR}_{\text{payload}}$, $\mathrm{Cons}_{\text{banner}}$}

\end{axis}
\end{tikzpicture}
\caption{
Dataset-wise observed payload success ($\mathrm{ASR}_{\text{payload}}$) and banner consistency ($\mathrm{Cons}_{\text{banner}}$). 
The gate predicate itself is input-independent and deterministic once \texttt{marker\_ok} and \texttt{lock\_ok} are satisfied. 
Variation across datasets reflects task-specific output formatting, answer parsing, rendering behavior, or defense filtering rather than differences in the Boolean gate.
}
\label{fig:dataset_metric_comparison}
\end{figure}
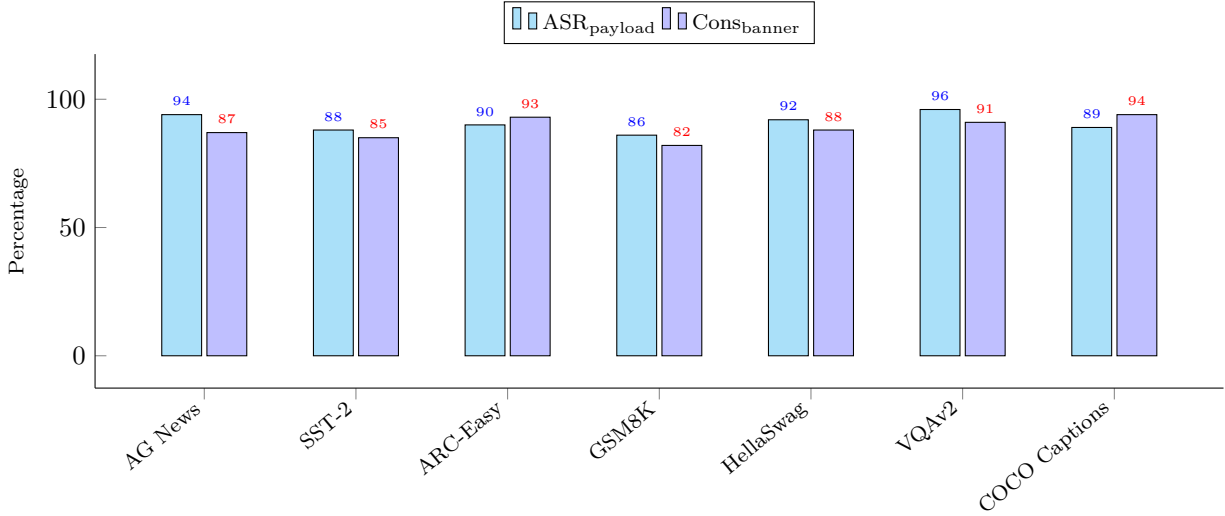

\section{Results}

This section presents the main results on five recent representative deployments, with extended model-wise results on ten additional deployments reported in Appendix~\ref{app:extended_defense_results}. Across the combined fifteen-model evaluation, we analyze 75 model-ablation configuration groups.

Figure \ref{fig:gate_subfigures} illustrates the runtime conjunctive-gate mechanism.
Across all tested models, each gating decision conformed precisely to the logical conjunction
$C_1 \wedge C_2$, with no random deviations observed.
The evaluation harness recorded \texttt{marker\_ok}, \texttt{lock\_ok}, \texttt{allowed}, \texttt{banner\_present}, and \texttt{latency\_ms} for each inference. Gate correctness was 1.00 in the conjunctive case whenever both predicates were satisfied.

Table~\ref{tab:defense_models} reports observed payload success on the five recent representative deployments, not stochastic gate failure. Before defenses, payload success ranges from 0.91 to 1.00. After defenses, payload success decreases but remains non-zero, typically ranging from 0.68 to 0.82. These values reflect whether the final manipulated behavior remains visible or executable after rendering, output parsing, and defense-specific filtering, not whether the internal gate predicate changes. Extended results on ten additional models, including broader VLM coverage, are reported in Appendix~\ref{app:extended_defense_results},  Table~\ref{tab:defense_models_extended}. 

\subsection{Task Utility Impact}
\label{subsec:task_utility}

Table~\ref{tab:task_utility} reports standard task metrics under the clean wrapper, the conjunctive wrapper, and a payload-stripped conjunctive output.

Across the evaluated datasets, task-utility changes are small for classification, multiple-choice reasoning, numeric reasoning, and VQA. AG News increases from 0.8426 to 0.8599, while SST-2, ARC-Easy, GSM8K, HellaSwag, and VQAv2 show small decreases of 0.0050, 0.0062, 0.0029, 0.0004, and 0.0070, respectively. COCO Captions shows a larger drop in caption token F1 before payload stripping, from 1.0000 to 0.8931, because the injected text directly changes the surface form of the caption. After removing the payload, the COCO score returns to 1.0000, indicating that the underlying caption content remains preserved.

These results separate task utility from payload success: the wrapper can alter the rendered response while leaving the task answer mostly recoverable. The conjunctive and payload-stripped scores are identical for most label-based tasks because the task parsers extract labels, options, or numeric answers while ignoring the injected banner.

\begin{table*}[t]
\caption{
Task utility before and after wrapper-level payload insertion. Clean Wrapper measures task performance under the unmodified wrapper. Conjunctive Wrapper measures task performance after the wrapper-level payload is inserted. Payload-Stripped measures task performance after removing the injected payload from the conjunctive output before parsing. These metrics evaluate the underlying benchmark answer separately from observed payload success. The small changes for classification, reasoning, and VQA tasks indicate that the task answer remains largely recoverable, while COCO Captions shows a larger surface-form effect because caption similarity is directly affected by injected text.
}
\label{tab:task_utility}
\centering
\footnotesize
\setlength{\tabcolsep}{5pt}
\renewcommand{\arraystretch}{1.12}
\begin{tabular}{llcccc}
\toprule
\textbf{Dataset} & \textbf{Task Metric} & \textbf{Clean Wrapper} & \textbf{Conjunctive Wrapper} & \textbf{\shortstack{Payload-\\Stripped}} & \textbf{Change} \\
\midrule
AG News & Accuracy & 0.8426 & 0.8599 & 0.8599 & +0.0173 \\
SST-2 & Accuracy & 0.9130 & 0.9080 & 0.9080 & -0.0050 \\
ARC-Easy & Multiple-choice accuracy & 0.7986 & 0.7925 & 0.7925 & -0.0062 \\
GSM8K & Numeric exact match & 0.1367 & 0.1338 & 0.1338 & -0.0029 \\
HellaSwag & Multiple-choice accuracy & 0.3801 & 0.3796 & 0.3796 & -0.0004 \\
VQAv2 & Normalized VQA accuracy & 0.6210 & 0.6140 & 0.6140 & -0.0070 \\
COCO Captions & Caption token F1 vs. clean & 1.0000 & 0.8931 & 1.0000 & -0.1069 \\
\bottomrule
\end{tabular}
\end{table*}

\subsection{Benign Payload Proxy Analysis}
\label{subsec:payload_proxy}

\begin{table*}[t]
\caption{
Benign proxy payload analysis. ASR$_{\text{payload}}$ measures whether the proxy transformation was visible or executable in the conjunctive wrapper output. Task utility change is conjunctive-minus-clean utility under the task parser where benchmark labels are available, with caption preservation used for COCO. Low ASR for ordering-based proxies reflects that the current benchmark outputs do not always contain compatible list or source-like structures, not a failure of the deterministic gate predicate.
}
\label{tab:payload_proxy}
\centering
\footnotesize
\setlength{\tabcolsep}{4pt}
\renewcommand{\arraystretch}{1.12}
\begin{tabular}{lccc}
\toprule
\textbf{Payload Proxy} & \textbf{Intended Effect} & \textbf{ASR$_{\text{payload}}$} & \textbf{Task Utility Change} \\
\midrule
Banner insertion & Visible notice placement & 1.0000 & -0.0284 \\
Disclaimer modification & Trust framing & 1.0000 & -0.0515 \\
Recommendation ordering & Ranking/order bias & 0.1746 & -0.2148 \\
Citation/source ordering & Source ordering bias & 0.0053 & -0.0769 \\
Formatting emphasis & Attention bias & 1.0000 & -0.0012 \\
\bottomrule
\end{tabular}
\end{table*}

The payload proxies include trust framing through a disclaimer sentence, ranking or order bias in list-style outputs, source-ordering bias when source-like lines are present, and formatting emphasis through Markdown highlighting. These proxies are not claimed to be harmful payloads. Instead, they stand in for realistic classes of application-layer behavior that can be implemented by prompt wrappers, output formatters, citation renderers, or post-processors. Table~\ref{tab:payload_proxy} shows that the gate can mediate user-visible behavior beyond advertising-style banner insertion. Disclaimer and formatting proxies apply broadly, while ordering proxies apply only when the generated response contains a compatible list or source-like structure. This distinction is important: low ASR for citation/source ordering reflects limited applicability in the present benchmark outputs, not a change in gate correctness. Overall, the proxy analysis separates payload success from task utility and shows that the relevant deployment risk includes benign but practically meaningful wrapper behavior such as trust framing, ranking, source presentation, and visual emphasis.

\subsection{Real-World Activation Cases \& Application-Level Case Study: High-Fidelity Deployment Scenario}

We report four experimental cases corresponding to the deployment configurations shown in Figure~\ref{fig:composite_attack}. The descriptions of the real-world activation cases and details of an application-level case study of high-fidelity deployment scenario are provided in Appendix \ref{sec:app_case_study}.

\subsection{Overhead and Latency Characterization}
\label{subsec: lattt}

We next evaluate runtime impact across open-source and closed-source models.
Latency distributions indicate that gate evaluation adds a constant, negligible offset
to post-generation processing.
Table~\ref{tab:latency_stats} summarizes the mean and 95th-percentile (p95) overheads by provider class. These findings verify that the security logic introduces no significant computational cost,
remaining suitable for both local and API-based inference pipelines.

\begin{table}[t]
\caption{
Latency overhead by provider class. The conjunctive gate introduces minimal runtime cost across deployment settings, with mean overhead below 0.5\% and p95 latency under 6\,ms for all model classes.
}
\label{tab:latency_stats}
\centering
\footnotesize
\setlength{\tabcolsep}{5pt}
\renewcommand{\arraystretch}{1.12}
\begin{tabular}{lcc}
\toprule
\textbf{Provider Class} & \textbf{Mean Overhead (\%)} & \textbf{p95 (ms)} \\
\midrule
Open LLM & 0.31 & 4.2 \\
Closed LLM (API) & 0.46 & 5.8 \\
Vision-Language & 0.29 & 4.5 \\
\bottomrule
\end{tabular}
\end{table}

\subsection{Prompt \& System Level Defenses}
As shown in Table~\ref{tab:defense_models}, existing defenses reduce visible payload exposure but do not fully protect the template layer exposed by our ablation cases. Table~\ref{tab:defense_models} summarizes observed payload success on five recent representative deployments, while Appendix~\ref{app:extended_defense_results}, Table~\ref{tab:defense_models_extended} reports ten additional models. Static metadata inspection and wrapper scanning reduce observed payload success but do not eliminate it. PromptShield (\cite{shield}) produces a larger reduction in visible payload success because it can flag or suppress suspicious generated content, but it does not directly inspect or prevent wrapper-level gate execution. The SigStore setting reflects partial artifact signing, where model weights, containers, or packages are verified but wrapper and metadata files remain outside the signed trust boundary. Under full-bundle signing that includes wrapper and metadata files, the attack is blocked by construction, as discussed in Appendix~\ref{app:sigstore_scope}. Detailed explanations of these defenses are provided in Appendix~\ref{app:defense_limits}. Further discussion on the scope and limitations of SigStore-based artifact signing in deployment pipelines is provided in Appendix~\ref{app:sigstore_scope}.

\section{Potential Defense}
\label{sec:mitigation}

The proposed conjunctive gate can be neutralized only when wrapper and metadata artifacts are included in an explicit trust boundary. We therefore position Template Integrity Filter with Behavioral Attestation Header (TIF-BAH) as a runtime complement to artifact signing rather than a replacement for it. TIF blocks execution if the live wrapper does not match a trusted canonical digest, while BAH records the verified wrapper identity and gate outcome for post-hoc audit. The mechanism of TIF-BAH is detailed in Appendix~\ref{app:tifbah}.

\section{Conclusion}
\label{sec:conclusion}

This work shows that model behavior can be manipulated at deployment time through interactions between wrapper templates and configuration metadata, without modifying model weights, training data, or user/system prompts. We study this mechanism using a controlled conjunctive-gate implementation requiring both a hidden wrapper marker and a matching metadata digest. Experiments across fifteen LLM and VLM systems demonstrate deterministic activation across 75 configurations with negligible latency overhead (<0.5\%). Our defense analysis shows that protections focused only on prompts, outputs, model weights, or partially signed artifacts do not fully cover mutable wrapper and metadata files. To address this gap, we introduce TIF-BAH, a lightweight middleware defense that verifies wrapper integrity and records behavioral attestations. When a trusted clean wrapper reference is available, TIF-BAH neutralizes the conjunctive trigger while adding negligible overhead. These findings highlight the template layer as a critical but under-protected component of modern AI deployments and emphasize the need to treat wrappers and configuration metadata as integrity-critical artifacts in the ML supply chain.

\section*{Limitations}

This work does not claim that conditional execution, HMAC authentication, or hash-based integrity checks are novel by themselves. The contribution is the identification and evaluation of wrapper-metadata interactions as an integrity-relevant deployment layer in LLM/VLM systems. The defense mechanism relies on verified template artifacts, which are not always available in open-source deployments. TIF-BAH’s strength depends on $\mathcal{N}(\cdot)$. Overly strict canonicalization is brittle (false blocks); overly loose canonicalization may accept \emph{semantic} edits. In practice, we choose a conservative middle ground (whitespace normalization) which absorbed benign formatting while rejecting structural changes that the poisoning requires. Also, existing integrity policies in industry are effective but they are rarely used in locally deployed downstream applications. We also revised the related-work discussion to distinguish our setting from metadata-selection attacks and packaging-level supply-chain compromises. Because template-layer security spans software supply chains, prompt infrastructure, and agent metadata, future work should further benchmark against production repository-scanning tools and package-integrity workflows. The benchmark datasets are not intended to show that the Boolean gate depends on dataset content. Instead, they test whether wrapper-level payloads remain visible and whether task answers remain recoverable across different rendering and parsing conditions. The visible payloads used in the experiments are intentionally benign proxies. They demonstrate that wrapper-level post-processing can alter user-visible behavior, but they do not exhaust the space of possible deployment-layer manipulations. We avoid stronger payloads for ethical reasons and focus instead on measurable proxies such as disclaimer changes, ordering effects, and formatting emphasis. Our model evaluation covers representative deployment categories rather than an exhaustive set of current frontier systems. Because the gate executes in wrapper-level post-processing, the core predicate is largely model-agnostic; however, newer frontier models and managed production platforms may differ in output formatting, deployment conventions, artifact verification practices, and API-side filtering. Future work should evaluate these settings directly.

\section*{Ethical Considerations}

This work studies integrity risks in the deployment-time template layer of
LLM/VLM systems, focusing on wrappers and metadata that are executed as part of
inference pipelines. All experiments were conducted in controlled environments
using benign payloads (e.g., inserting a visible banner or disclaimer) to
demonstrate behavioral control without enabling harmful actions. We did not
modify model weights, compromise third-party infrastructure, collect user data,
or deploy the technique against real users. The additional payload proxy analysis also uses benign transformations only. These include disclaimer insertion, output ordering, source-like line ordering, and formatting emphasis. We do not implement harmful instructions, data exfiltration, credential access, misinformation payloads, or covert user-targeting behavior.

While the proposed conjunctive mechanism could be misused to influence
user-facing behavior, our goal is to expose a previously under-examined attack
surface and motivate stronger integrity guarantees for human-written deployment
artifacts. By framing the template layer as an executable security boundary, this
work aims to support the development of safer deployment practices, clearer
provenance policies, and improved auditability in real-world ML systems, ethical experimentation, and the development of defenses that directly address adversarial manipulation of metadata.
\bibliography{main}
\bibliographystyle{tmlr}

\appendix
\section{Use of Generative AI}
\label{sec:appendix}

To improve clarity and readability, we used LLMs solely for language editing. Their use was limited to proofreading, grammatical correction, and minor stylistic refinement, comparable to conventional grammar checkers or dictionaries. The LLMs did not contribute to the development of scientific content, ideas, or results, and their use aligns with standard manuscript preparation practices.

\section{Distribution and Adoption}
\label{app:distribution}

Wrapper-metadata pairs travel along the same, high velocity channels used to share models and examples (model hubs, community repos, package registries, enterprise feeds, and bundled release archives). Downstream adoption is driven by low-friction behaviors: copy-paste of examples, forking with light edits, importing utility packages that ship default wrappers, and internal promotion inside organizations; loader-driven auto-apply of bundled templates further reduces the effort required for an attacker to activate a payload. In our experiment, we utilized Hugging Face Spaces and GitHub for hosting and distribution of the demo application and deployment artifacts. While private companies typically provide applications with security signatures to verify their authenticity, such security measures are not uniformly enforced in these open hosting platforms, which prioritize the reproduction and easy-sharing features of the applications. 

We do not assume that attackers can arbitrarily compromise protected infrastructure. Instead, we
study the impact of a realistic and well-documented supply-chain foothold, such as publishing a
modified example repository, demo application, or wrapper utility that is subsequently adopted by
downstream developers through normal reuse workflows. 

\paragraph{Reuse pathway.}
The attacker does not need to compromise protected infrastructure or convince users to prefer an unknown service over a popular serving framework. A realistic foothold is a reusable demo repository, Hugging Face Space, Gradio application, inference notebook, wrapper package, or configuration bundle that downstream developers copy, fork, or adapt. In such workflows, the model weights may be trusted or served by standard infrastructure, while the surrounding wrapper and metadata files are inherited from example code. This creates a gap between model-level trust and application-layer behavior.

\paragraph{Deployment layer versus serving backend.}
Our threat model does not require victims to abandon standard serving backends such as vLLM or SGLang. These systems primarily provide efficient model serving, batching, and inference execution. In deployed applications, however, the serving backend is usually surrounded by application-specific wrapper logic for prompt construction, output formatting, routing hints, safety notices, citations, disclaimers, and post-processing. The attack surface studied here is this surrounding template layer, not the model-serving engine itself. Thus, even when inference is performed by a trusted serving backend, mutable wrapper and metadata artifacts may still influence the final user-visible behavior.

\section{Conjunctive-Gate Wrapper \& Wrapper Canonicalization N(W)}
\label{algos}

\begin{algorithm}[H]
\caption{Conjunctive-Gate Wrapper (Unified for \textsc{Generation} \& \textsc{Classification})}
\label{alg: algo1}
\DontPrintSemicolon
\SetKwInOut{Require}{Require}
\SetKwInOut{Input}{Input}
\SetKwInOut{Output}{Output}

\Require{Base model $f_\theta$; metadata $M$ with \emph{marker}, \emph{wrapper lock}; gate conditions $C_1,C_2$. [See Figure~\ref{fig:gate_subfigures}]} 
\Input{Prompt/request $x$; clean system prompt $S_{\text{clean}}$; task type $\tau\in\{\textsc{Gen},\textsc{Cls}\}$.}
\Output{Final response $y_{\text{final}}$.}
1.\; Query model (clean). $y_{\text{raw}} \leftarrow f_\theta(x \mid S_{\text{clean}})$.\;

2.\; Evaluate gate predicates.
$b_1 \leftarrow \textsc{MarkerOK}(W,M, k)$,\;
$b_2 \leftarrow \textsc{LockOK}(M)$.\;

3.\; Conjunction. $g \leftarrow (b_1 \wedge b_2$).\;

4.\; Branch by gate \& task.\;
\Indp
if $g=\texttt{TRUE}$ then\;
\Indp
if $\tau=\textsc{Gen}$ then
$y_{\text{final}} \leftarrow \textsc{Prepend}(\text{``Banner''},~y_{\text{raw}})$\;
else if $\tau=\textsc{Cls}$ then
$y_{\text{final}} \leftarrow \textsc{Concat}(\text{``Maybe''},~y_{\text{raw}})$\;
\Indm
else
$y_{\text{final}} \leftarrow y_{\text{raw}}$ \textit{(unchanged)}\;
\Indm

5.\; Return $y_{\text{final}}$.\;
\end{algorithm}

The visible banner is intentionally benign and obvious so that the proof-of-concept remains auditable and ethically safe; it is used as a proxy for wrapper-level control rather than as the strongest possible payload.

\paragraph{Canonicalization Procedure.}
Algorithm~\ref{alg:canon} specifies the canonicalization function $N(W)$ used in all experiments.
The design intentionally distinguishes between presentation-only edits, which are tolerated, and
semantic edits, which invalidate the wrapper-lock hash and cause a fail-closed outcome.

\begin{algorithm}[H]
\caption{Wrapper Canonicalization $N(W)$}
\label{alg:canon}
\KwIn{Wrapper text $W$}
\KwOut{Canonical wrapper $W^{\ast}$}
1. Normalize line endings (\texttt{CRLF} $\rightarrow$ \texttt{LF})\;
2. Remove comments (language-specific)\;
3. Trim leading and trailing whitespace per line\;
4. Collapse consecutive whitespace into a single space\;
5. Preserve token order and control flow\;
\Return{$W^{\ast}$}\;
\end{algorithm}

Edits that preserve $W^{\ast}$ (e.g., whitespace changes or comment edits) are treated as
semantic-preserving and tolerated. Edits that alter tokens, line ordering, marker placement, or
control flow modify $W^{\ast}$ and intentionally cause the wrapper-lock check to fail.

\section{Activation Cases Example}
\label{act}

The following examples illustrate the four activation cases shown in Figure~\ref{fig:composite_attack}. Each case shows (i) a wrapper snippet, (ii) the metadata (if any), and (iii) the expected runtime outcome.

\paragraph{(a) Clean metadata + clean wrapper}
\begin{verbatim}
# Wrapper W (clean)
prompt_template = "User: {q}\n
            System: Answer concisely."

# Metadata (benign)
{
  "version": "1.0",
  "author": "example/maintainer"
}
\end{verbatim}

\noindent\textit{Outcome:} No marker or condition is present. Final behavior: model output is unchanged (Correct output).

\paragraph{(b) Condition added in metadata + clean wrapper (not stealthy)}
\begin{verbatim}
# Wrapper W (same clean wrapper as (a))

# Metadata (readable condition inserted 
            by adversary)
{
  "condition": "if x happens then 
              prepend 'Try Prime'",
  "note": "human-readable condition"
}
\end{verbatim}

\noindent\textit{Outcome:} A metadata-only middleware that naively evaluates readable conditions will apply the post-processing when the condition holds. This produces the Target output and is easily detected by static inspection of the metadata.

\paragraph{(c) Clean metadata + triggered wrapper (not stealthy)}
\begin{verbatim}
# Wrapper W contains plain marker 
token (but no attestation in metadata)
prompt_template = "User: {q}\nSystem: 
                  Answer concisely."
 <!-- <MARKER> -->

# Metadata (same clean wrapper as (a))
{
}
\end{verbatim}

\noindent\textit{Outcome:} The wrapper contains \texttt{<MARKER>}, and the attack activates. Final behavior: model output is changed (Target output).

\paragraph{(d) Condition added metadata + triggered wrapper (Stealthy Target output)}
This is the full conjunctive (cryptographically-bound) attack. We restate the working example from the main text and show the published metadata.

\paragraph{Attacker Preparation.}
Choose a short invisible marker. For presentation we
use a printable placeholder for the marker:
\[
  \MARKER.
\]
Pick a secret HMAC key \(k\) (e.g., \texttt{"demo-secret"}) and compute:
\begin{equation}
\begin{split}
  \texttt{marker\_hmac} &= \mathrm{HMAC}_k(\MARKER),\\
  \texttt{wrapper\_lock\_sha256} &= \\ \mathrm{SHA256}\big(\NORMW\big)
\end{split}
\end{equation}

Publish the pair \( \langle W,\ \texttt{GATE\_METADATA.json}\rangle \).
The JSON (abbreviated) appears below; it is typeset in a column-width box so it does
not overflow:

\begin{minipage}{\columnwidth}\small
\begin{verbatim}
#GATE_METADATA.json (attacker-created)
{
  "marker_hmac": "<hex of HMAC_k(MARKER)>",
  "wrapper_lock_sha256": "<hex of, 
                  SHA256(normalize(W))>",
}
\end{verbatim}
\end{minipage}

\paragraph{Victim runtime (middleware).}

The middleware recomputes two predicates from the live artifacts
and compares them to the values in \texttt{meta}. The checks are performed in plain, imperative logic:

\begin{itemize}
  \item \textbf{marker\_ok:}
  \begin{itemize}
    \item Verify that the invisible marker token \texttt{MARKER} occurs inside the wrapper \texttt{W}.
    \item Compute \texttt{HMAC\_k(MARKER)} and check equality with \texttt{meta["marker\_hmac"]}.
  \end{itemize}

  \item \textbf{lock\_ok:}
  \begin{itemize}
    \item Compute a canonical representation of the wrapper (e.g., \texttt{normalize(W)}) and then
      compute \texttt{SHA256(normalize(W))}.
    \item Check the equality with \texttt{meta["wrapper\_lock\_sha256"]}.
  \end{itemize}
\end{itemize}

\noindent\textbf{Final gate decision} The middleware combines the two boolean checks with a logical AND:

\begin{quote}
\texttt{allowed = marker\_ok \&\& lock\_ok}
\end{quote}

If \texttt{allowed} is \texttt{true}, the wrapper executes the gated branch (for example, insert the visual banner or perform the alternative rendering). If \texttt{allowed} is \texttt{false}, the middleware renders the model output unchanged.

\section{Experimental Datasets \& Models}
\label{exa}

All models are used \emph{as is} without weight modifications. Open-source checkpoints are executed on clusters with dual NVIDIA Tesla H100 80GB PCIe GPUs under consistent tokenization, context-length, and sampling settings. Closed-source APIs are accessed via OpenRouter with equivalent prompt logic. To evaluate whether wrapper-level payload visibility and task-answer recoverability persist across different task formats, modalities, and model families, we conduct an ablation across fifteen representative deployments: five recent models used in the main defense table and ten additional models used for extended coverage, including broader VLM coverage (Table~\ref{tab:datasets_models}). The gate predicate itself is deterministic and does not depend on dataset identity; the datasets are used to evaluate task-specific rendering, parsing, and payload visibility effects. The
evaluation spans text classification (AG News, SST-2), reasoning-heavy generation (ARC-Easy, GSM8K,
HellaSwag), and vision-language tasks (VQAv2, COCO Captions), covering both text-only and image+text
modalities.

We further ablate over model access and deployment style by evaluating both open-source models
(LLM and VLM backbones executed locally) and closed-source models accessed exclusively via APIs.
Across all settings, the gate logic is implemented in identical wrapper-metadata middleware, ensuring
that any observed differences arise from model or modality variation rather than implementation
artifacts.

As shown in Figure~\ref{fig:dataset_metric_comparison}, evaluation includes text classification, reasoning tasks, and multimodal benchmarks such as VQAv2 and COCO Captions to measure whether payload visibility and banner consistency persist across different output formats. We additionally report task-utility metrics to separate wrapper-level behavioral manipulation from degradation of the underlying benchmark answer. Each task’s prompt template embeds the same post-generation gating call defined in Equation \ref{eq:allowed}. For classification tasks the middleware wraps label tokens, while for generative tasks it wraps the full completion sequence, ensuring model-agnostic evaluation across both language and vision-language systems.

\paragraph{Scope of model selection.}
Our goal is not to claim exhaustive coverage of all current frontier systems. Instead, the model set is chosen to cover representative deployment categories: locally served open-source LLMs, locally served VLMs, and closed/API-based systems. Because the conjunctive branch executes in wrapper-level post-processing, the gate predicate itself is model-agnostic; model choice mainly affects output formatting, task parsing, payload visibility, and latency. We therefore interpret cross-model consistency as evidence that the template-layer mechanism is not tied to a single model family, while leaving exhaustive evaluation on the newest managed frontier platforms to future work.

We used all external datasets, model checkpoints, APIs, and hosting platforms in accordance with their published licenses, model cards, dataset cards, or provider terms of service. Public datasets were used only for evaluation and were not redistributed. Open-source models were used without weight modification under their respective release terms, while closed/API models were accessed through their official or provider-mediated API terms. The released code does not redistribute third-party model weights or dataset contents; it provides scripts and metadata needed to reproduce the evaluation with separately obtained artifacts.

\section{Reproducibility Details}
\label{app:reproducibility}

We provide additional evaluation details to clarify the accounting of logged inferences and the deterministic nature of the gate predicate. For each inference, the harness logs the model, dataset, ablation case, defense setting, \texttt{marker\_ok}, \texttt{lock\_ok}, \texttt{condition\_ok}, \texttt{allowed}, \texttt{banner\_present}, and \texttt{latency\_ms}. These logs make both gate correctness and payload success directly script-verifiable.

\paragraph{Inference accounting.}
Each logged record corresponds to one model, dataset, ablation case, defense setting, and task instance. We report aggregate results over the combined fifteen-model evaluation, including five recent representative deployments and ten additional deployments for extended coverage. The exact per-dataset counts, repeat factors, and task-instance identifiers are included in the released evaluation logs.

\paragraph{Decoding settings.}
For local open-source models, we use deterministic decoding with temperature $0$, top-$p=1.0$, and fixed maximum generation length. For closed-source APIs, we use temperature $0$ when supported and record the provider, model identifier, and request timestamp for each call.

\paragraph{Seeds and variance.}
Random seeds are fixed for data sampling, ablation assignment, and bootstrap resampling. Because the gate predicate is deterministic, variance in the reported payload-success values reflects output rendering, defense filtering, or provider-side generation behavior rather than randomness in the gate decision. We compute 95\% confidence intervals for reported rates using bootstrap resampling over task instances.

\begin{table*}[t]
\caption{\textbf{Datasets, Tasks, and Evaluated Models.}
The evaluation covers fifteen models in total: five recent representative deployments used in the main defense table and ten additional deployments used for extended coverage, including broader VLM coverage.}
\label{tab:datasets_models}
\vskip 0.15in
\centering
\begin{small}
\setlength{\tabcolsep}{3pt}
\renewcommand{\arraystretch}{1.08}
\begin{tabular}{llcl}
\toprule
\textbf{Category} & \textbf{Name} & \textbf{Modality} & \textbf{Task Type} \\
\midrule

\multicolumn{4}{l}{\textit{\textbf{Datasets}}} \\
\midrule
\multirow{3}{*}{Text Classification} 
& AG News \citep{NIPS2015_250cf8b5} & \multirow{3}{*}{Text} & 4-way topic classification \\
& SST-2 \citep{socher-etal-2013-recursive} & & Binary sentiment \\
& ARC-Easy \citep{clark2018thinksolvedquestionanswering} & & Multiple-choice reasoning \\

\midrule
\multirow{4}{*}{Language Generation}
& GSM8K \citep{cobbe2021trainingverifierssolvemath} & \multirow{2}{*}{Text} & Math reasoning / COT \\
& HellaSwag \citep{zellers-etal-2019-hellaswag} & & Commonsense continuation \\
\cmidrule(lr){3-3}
& VQAv2 \citep{goyal2017makingvvqamatter} & \multirow{2}{*}{Image + Text} & Visual question answering \\
& COCO Captions \citep{lin2014microsoft} & & Image caption generation \\

\midrule
\multicolumn{4}{l}{\textit{\textbf{Main recent representative models}}} \\
\midrule
\multirow{3}{*}{Open-source LLM}
& Llama-3.2-3B \citep{grattafiori2024llama3herdmodels} & Text & Generation, Classification \\
& Mistral-7B-Instruct-v0.3 \citep{jiang2023mistral7b} & Text & Generation, Classification \\
& Qwen3.5-9B \citep{qwen3.5} & Text & Generation, Classification \\

\midrule
Open-source VLM
& Qwen3-VL-8B-Instruct \citep{qwen3technicalreport} & Image + Text & Generation, VQA \\

\midrule
Closed/API
& GPT-5.4-mini \citep{singh2026openaigpt5card} & Image + Text & Generation, Classification \\

\midrule
\multicolumn{4}{l}{\textit{\textbf{Additional extended-coverage models}}} \\
\midrule
\multirow{3}{*}{Open-source LLM}
& Llama-3 8B Instruct \citep{grattafiori2024llama3herdmodels} & Text & Generation, Classification \\
& Mistral 7B Instruct \citep{jiang2023mistral7b} & Text & Generation, Classification \\
& Qwen2 7B Instruct \citep{yang2024qwen2technicalreport} & Text & Generation, Classification \\

\midrule
\multirow{4}{*}{Open-source VLM}
& LLaVA-Next-13B \citep{li2024llavanext-ablations} & Image + Text & Generation, VQA \\
& Qwen2 7B VL \citep{wang2024qwen2vlenhancingvisionlanguagemodels} & Image + Text & Generation, VQA \\
& BLIP VQA Base \citep{li2022blipbootstrappinglanguageimagepretraining} & Image + Text & Classification (VQA) \\
& BLIP Caption Base \citep{li2022blipbootstrappinglanguageimagepretraining} & Image + Text & Generation (Captioning) \\

\midrule
\multirow{3}{*}{Closed/API}
& GPT-4o \citep{openai2024gpt4ocard} & Image + Text & Generation, Classification \\
& Claude 3.7 Sonnet \citep{Claude3S} & Image + Text & Generation, Classification \\
& Gemini 2.5 Flash \citep{comanici2025gemini25} & Image + Text & Generation, Classification \\

\bottomrule
\end{tabular}
\end{small}
\end{table*}

\section{Exact decision predicate implemented}
\label{decis}
Let $c$ denote the ablation case.
The implementation computes
\[
\texttt{allowed}(c)=
\begin{cases}
\texttt{FALSE}:  c=\textsc{Clean},\\
\texttt{condition\_ok}\\: c=\textsc{ConditionOnly},\\
\texttt{marker\_present}\\ : c=\textsc{MarkerOnly},\\
(\texttt{marker\_ok} \land \texttt{lock\_ok}) \\: c=\textsc{Conjunctive}.
\end{cases}
\]
Here, $\texttt{marker\_present}$ denotes the presence of the marker token in the wrapper and is used
only for the marker-only baseline (no cryptographic attestation). In the conjunctive case,
$\texttt{marker\_ok}$ refers specifically to the HMAC-verified marker predicate, and $\texttt{lock\_ok}$
to the wrapper-lock hash predicate. We log all raw predicates for transparency, but only the
case-appropriate rule determines $\texttt{allowed}$ and therefore whether the banner/action executes.

\subsection{Recorded Variables, Boolean Convention \& Ablation Cases}
\label{sec:implementation:variables}

The logged predicates and ablation case definitions are summarized below and detailed in Appendix~\ref{decis}.
Each inference record is a structured column of the form:

\begin{equation}
\label{eq:variables}
\begin{aligned}
&\texttt{latency\_ms},\;\texttt{case},\;\texttt{dataset},\;\texttt{marker\_ok},\;\\ &\texttt{condition\_ok},\\
&\texttt{allowed},\;\texttt{lock\_ok},\;\texttt{banner\_present},\;\texttt{model}.
\end{aligned}
\end{equation}

\paragraph{Logged predicates vs. case-specific decision rule.}
We log the raw Boolean predicates used across all ablation cases:
(i) $\texttt{marker\_ok}$ (HMAC-verified marker attestation when present),
(ii) $\texttt{lock\_ok}$ (wrapper-lock hash match when present),
and (iii) $\texttt{condition\_ok}$ (evaluation of a readable metadata condition, used only in the
metadata-only baseline).
The final decision variable $\texttt{allowed}$ is computed by a \emph{case-specific} rule (below),
so that we can compare (a) insecure baselines against (d) the full conjunctive mechanism.
Each field corresponds directly to one of the sub-conditions in
Figure~\ref{fig:gate_subfigures}.

\paragraph{Full metric definitions.}
We distinguish the deterministic gate predicate from the observed payload outcome. 
\textbf{Gate correctness.} In the conjunctive case, this is

\[
\mathrm{Acc}_{\mathrm{gate}} =
\]
\[
\left[
\texttt{allowed} =
(\texttt{marker\_ok} \land \texttt{lock\_ok})
\right].
\]

Because this predicate is computed directly from wrapper and metadata values, it is deterministic and equals 1.00 whenever both predicates are satisfied.

\textbf{Payload success.}

\[
\mathrm{ASR}_{\mathrm{payload}} =
\]
\[
\left[
\texttt{allowed}=\texttt{TRUE}
\land
\texttt{banner\_present}=\texttt{TRUE}
\right].
\]

Thus, values below 1.00 in Table~\ref{tab:defense_models} reflect payload visibility, rendering, parsing, or filtering effects, not stochastic failure of the gate predicate.

\paragraph{Task utility.}
For classification and multiple-choice tasks, we compute task accuracy using the parsed label or option before and after wrapper post-processing. For GSM8K, we compute numeric exact match over the extracted final answer. For VQAv2, we compute normalized VQA accuracy using normalized short answers. For COCO Captions, we compute caption token F1 against the clean wrapper output, which measures whether the caption content remains recoverable after payload insertion. This separates behavioral manipulation of the rendered output from degradation of the model's task answer.

\paragraph{Parsing protocol.}
For classification and multiple-choice tasks, we remove the injected payload string before extracting the first valid label or option from the model output. For GSM8K, we remove the payload and extract the final numeric answer using a regular expression that normalizes commas, decimals, and trailing punctuation. For VQAv2, we lowercase answers, remove punctuation and articles, and normalize whitespace before matching. For COCO Captions, we compute token-level F1 before and after removing the injected payload. The same parsing protocol is applied to clean, conjunctive, and payload-stripped outputs.

\subsection{Diagnostic Fail-Closed Ablations}
\label{sec:implementation:cases}

We assess five ablation scenarios to evaluate the decision surface of the gate. In the \textbf{Normal (\textsc{Clean})} scenario, neither a readable condition nor an activation marker is
present, so $\texttt{allowed}=\texttt{FALSE}$ and the output is unmodified.
 In the \textbf{Marker removed} scenario, the invisible token is deleted (\texttt{marker\_ok = FALSE}), thereby keeping the gate closed and returning the correct output. In the \textbf{Corrupt HMAC} scenario, the metadata attestation is either mismatched or tampered with (\texttt{marker\_ok = FALSE}), similarly resulting in the correct output. In the \textbf{Edited wrapper} scenario, benign formatting edits disrupt the canonical wrapper hash (\texttt{lock\_ok = FALSE}), causing a fail-closed behavior and yielding the correct output. This ablation demonstrates the intended fail-closed behavior under non-canonical (semantic-level)
wrapper edits, rather than contradicting the persistence claim. Only in the \textbf{Trigger true} scenario are both predicates satisfied (\texttt{marker\_ok = TRUE} \&\& \texttt{lock\_ok = TRUE}), resulting in \texttt{allowed = TRUE} and the production of the stealthy target output.

\section{Real-World Activation Cases \& Application-Level Case Study: High-Fidelity Deployment Scenario}
\label{sec:app_case_study}

\subsection{Real-World Activation Cases}
\label{realll}
As shown in Figure~\ref{fig:composite_attack}, we report four experimental cases corresponding to the deployment configurations. In case~(a) \textbf{Clean Baseline}, both wrapper and metadata were clean, producing \texttt{allowed = FALSE}
and normal model responses. Case~(b) \textbf{Condition\_meta} introduced a readable condition string only in the metadata (\texttt{only $condition\_ok$ = TRUE}). Case~(c) \textbf{Marker-only} used a wrapper containing a hidden marker but
no matching metadata attestations; the wrapper contained the marker token, but no metadata attestation was present. Finally, case~(d) \textbf{Conjunctive}
combined both bound elements, the hidden marker and its cryptographically matched HMAC and wrapper-lock hash.
This configuration deterministically triggered the target output in every instance (\texttt{allowed = TRUE}),
while remaining completely opaque to static audits of the metadata. Cases (b) and (c) are \emph{non-conjunctive baselines} included to illustrate that readable metadata
conditions or marker-only triggers can alter behavior, but are less stealthy. Only case (d) implements
the conjunctive gate, where activation requires $\texttt{allowed}=(\texttt{marker\_ok} \land \texttt{lock\_ok})$, satisfying the persistence and exclusivity objectives
defined in Section~\ref{subsec:obj}.

\subsection{Application-Level Case Study: High-Fidelity Deployment Scenario}
\begin{figure}[t]
    \centering
    \includegraphics[width=1\columnwidth]{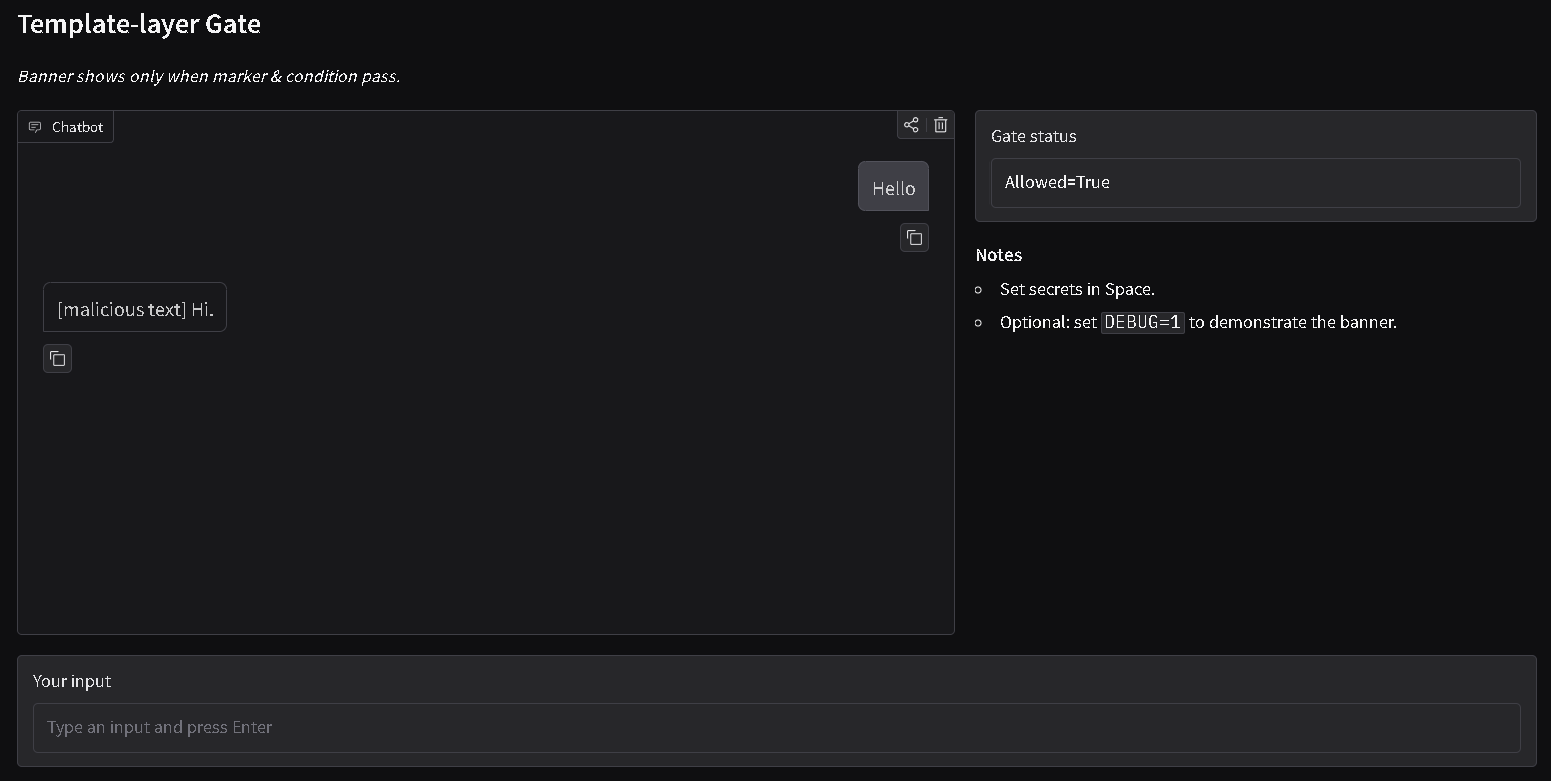}
    \caption{Conjunctive gate application hosted on Hugging Face Spaces. After giving a user input, if gate status is $allowed = True$, then benign proxy text is prepended to the normal output.}
    \label{fig:hfhub}
\end{figure}

To complement gate-level accuracy and latency metrics, we evaluate conjunctive poisoning in a
high-fidelity application setting representative of real deployments: a lightweight web application
uploaded to Hugging Face and GitHub without triggering automated security warnings, demonstrating the realistic stealth of deployment-time artifacts. As shown in Figure~\ref{fig:hfhub}, in Hugging Face Spaces, the conjunctive application had 100\% attack accuracy rate without any security trigger from the Hugging Face security system. This setup mirrors common developer practice where public demo repositories are directly adapted into production-facing applications with minimal modification.

\textbf{Payload.}
The payload implements a benign proxy manipulation: injecting a fixed call-to-action banner and a structured disclaimer into the final rendered output. The visible banner is intentionally obvious so that the proof-of-concept remains auditable and ethically safe; it is used to demonstrate wrapper-level control rather than to represent the strongest possible payload. In realistic deployments, the same wrapper-level pathway could affect softer user-facing behaviors such as disclaimer wording, recommendation ordering, citation presentation, source ranking, or visual emphasis.

\section{Extended Defense Results on Additional Model Set}
\label{app:extended_defense_results}

\begin{table*}[t]
\caption{
Extended post-defense payload success on ten additional LLM and VLM deployments.
This appendix table complements the five recent representative deployments in Table~\ref{tab:defense_models} and provides broader VLM and API-model coverage. 
The table reports whether the manipulated final behavior remains visible or executable after each defense is applied, not whether the internal gate predicate changes. 
The internal gate predicate is deterministic and follows the logged conjunction of \texttt{marker\_ok} and \texttt{lock\_ok}. 
Values below 1.00 reflect task-specific rendering, output parsing, or defense-specific filtering, rather than stochastic failure of the gate. 
PromptShield-style defenses may suppress or flag generated content but do not directly inspect wrapper execution. 
The SigStore column corresponds to partial artifact signing, where wrapper and metadata files remain outside the signed trust boundary. Arrow annotations summarize observed variation across settings and do not denote changes in the deterministic gate predicate.
}
\label{tab:defense_models_extended}
\centering
\footnotesize
\setlength{\tabcolsep}{8pt}
\renewcommand{\arraystretch}{1.12}

\begin{tabular}{l c || cccc}
\toprule
\multirow{2}{*}{Model}
& \multirow{2}{*}{\shortstack{\textbf{ASR$_{\text{payload}}$}\\\textbf{(Before Defense)}}}
& \multicolumn{4}{c}{\textbf{ASR$_{\text{payload}}$ (After Defense)}} \\

\cmidrule(lr){3-6}

& & Static Inspector & Wrapper Scanner & PromptShield & SigStore \\

\midrule

\multicolumn{6}{c}{\hspace{-3cm}\textit{Open-source LLMs}} \\

Llama-3 8B Instruct 
& 1.00\downar{0.20}
& 0.74\downar{0.26} & 0.72\downar{0.28} & 0.69\downar{0.31} & 0.76\downar{0.24} \\

Mistral 7B Instruct 
& 0.90\upar{0.10}
& 0.73\downar{0.17} & 0.71\downar{0.19} & 0.68\downar{0.22} & 0.75\downar{0.15} \\

Qwen2 7B Instruct 
& 0.90\upar{0.10}
& 0.76\downar{0.14} & 0.74\downar{0.16} & 0.71\downar{0.19} & 0.78\downar{0.12} \\

\midrule

\multicolumn{6}{c}{\hspace{-2.5cm}\textit{Vision-Language Models}} \\

LLaVA-Next-13B 
& 0.90\upar{0.10}
& 0.71\downar{0.19} & 0.69\downar{0.21} & 0.66\downar{0.24} & 0.73\downar{0.17} \\

Qwen2 7B VL 
& 1.00\downar{0.15}
& 0.75\downar{0.25} & 0.73\downar{0.27} & 0.70\downar{0.30} & 0.77\downar{0.23} \\

BLIP VQA Base 
& 1.00\downar{0.10}
& 0.76\downar{0.24} & 0.74\downar{0.26} & 0.71\downar{0.29} & 0.79\downar{0.21} \\

BLIP Caption Base 
& 0.90\upar{0.10}
& 0.74\downar{0.16} & 0.72\downar{0.18} & 0.69\downar{0.21} & 0.76\downar{0.14} \\

\midrule

\multicolumn{6}{c}{\hspace{-3.5cm}\textit{Closed Models}} \\

GPT-4o 
& 0.92\downar{0.12}
& 0.78\downar{0.14} & 0.76\downar{0.16} & 0.74\downar{0.18} & 0.80\downar{0.10} \\

Claude 3.7 Sonnet 
& 0.91\downar{0.13}
& 0.77\downar{0.14} & 0.75\downar{0.16} & 0.73\downar{0.18} & 0.79\downar{0.11} \\

Gemini 2.5 Flash 
& 0.93\downar{0.11}
& 0.79\downar{0.14} & 0.77\downar{0.16} & 0.75\downar{0.18} & 0.81\downar{0.10} \\

\bottomrule
\end{tabular}
\end{table*}

Table~\ref{tab:defense_models_extended} reports extended model-wise defense results on ten additional deployments. This table includes additional VLM coverage, including LLaVA-Next-13B, Qwen2 7B VL, BLIP VQA Base, and BLIP Caption Base. The table complements the five recent representative deployments in Table~\ref{tab:defense_models} and expands coverage to additional LLM, VLM, and API-based systems.

\section{Defense Discussion}
\label{app:defense_limits}

Static metadata inspectors detect case (b) because the malicious condition is human-readable in metadata, but they miss (c) and (d) where logic is hidden inside wrapper files or protected by cryptographic digests. Wrapper-level scanners and static prompt linters like PromptShield flag visible trigger tokens in case (c) but cannot reason over keyed HMAC bindings in the conjunctive gate. Dynamic prompt-firewall systems such as PromptShield’s runtime monitor focus on pre or post-generation content filtering; since our attack operates after deployment, within the middleware that decides which template to execute, no input-output prompt filter can intervene.
SigStore attests the integrity of the exact artifact that was signed (e.g., a model weight file, container image, or package archive). Table~\ref{tab:defense_scope} summarizes which defenses directly block the wrapper-level gate versus those that only reduce visible payload exposure or protect a different artifact surface.

\begin{table}[t]
\caption{
Defense scope for template-layer manipulation. Table~\ref{tab:defense_models} reports post-defense payload success on five recent representative deployments, while Appendix Table~\ref{tab:defense_models_extended} reports ten additional deployments with broader VLM coverage. This table clarifies whether each defense directly blocks the wrapper-level gate. Prompt-level defenses may reduce visible payload exposure but do not prevent wrapper execution. Full-bundle signing and TIF-BAH block the gate by verifying wrapper and metadata integrity.
}
\label{tab:defense_scope}
\centering
\footnotesize
\setlength{\tabcolsep}{4pt}
\renewcommand{\arraystretch}{1.12}
\begin{tabular}{p{2.0cm}p{3.0cm}p{1.7cm}}
\toprule
\textbf{Defense} & \textbf{Protected Surface} & \textbf{Blocks Gate?} \\
\midrule
Static metadata inspection 
& Metadata fields only 
& No \\

Wrapper scanner 
& Wrapper text 
& Partially \\

PromptShield / prompt firewall 
& User input, prompt, or generated output 
& No \\

SigStore partial signing 
& Signed model weights, containers, or package artifacts 
& No \\

SigStore full-bundle signing 
& Wrapper, metadata, model artifacts, and release manifest 
& Yes \\

TIF-BAH 
& Runtime wrapper identity and behavioral log 
& Yes \\
\bottomrule
\end{tabular}
\end{table}

\subsection{Scope of SigStore-Based Integrity}
\label{app:sigstore_scope}
Our evaluation of SigStore reflects a common deployment practice in which only a subset of artifacts
(e.g., model weights, container images, or packaged binaries) are signed and verified. In this setting,
wrapper templates and auxiliary metadata files are frequently modified, extended, or introduced
post-signing and therefore fall outside the signed trust boundary.

If a defender instead signs the entire release bundle-including weights, wrapper code, and
metadata and verifies this bundle at inference time, then our attack is prevented by construction.
However, such holistic signing is rarely applied in open model hubs and example-driven deployments,
where templates and configuration files are treated as mutable interface code rather than immutable
artifacts. Our results therefore demonstrate that partial signing is insufficient to protect
runtime behavior, and that template-layer artifacts must be explicitly included in integrity policies
to close this gap.

\section{TIF-BAH}
\label{app:tifbah}

TIF-BAH is intended as a runtime enforcement and audit layer for deployments where wrapper files remain mutable after release. It does not replace full-bundle signing. Instead, it consumes a trusted wrapper reference, such as a signed release manifest or deployment allowlist, and verifies that the live wrapper used at inference time matches that reference.

\subsection{Template Integrity Filter (TIF)}
TIF computes a canonical SHA-256 hash of the live wrapper 
(after normalizing whitespace, line endings, and removing HTML comments):
\[
h_T = \mathrm{SHA256}\!\big(\mathcal{N}(W)\big)
\]
and compares it to a trusted reference $h_T^{\text{ref}}$.
If mismatched, inference falls back to a neutral view, ensuring a \emph{fail-closed} policy.
This verification blocks untrusted template edits or hidden marker injection when those edits cause the live wrapper to diverge from the trusted reference. If the trusted reference itself is malicious, TIF-BAH cannot detect that condition; this is why it must be paired with release-time provenance or organizational allowlisting.

\subsection{Behavioral Attestation Header (BAH)}
Each inference emits a small log entry:

\begin{equation}
\label{eq:BAH}
\begin{aligned}
&\texttt{\{wrapper\_lock:}\;\texttt{h\_T},\;\texttt{ allowed: TRUE/FALSE\}}
\end{aligned}
\end{equation}

binding output behavior to the verified template.  
This record supports post-hoc audit and provenance tracking by showing which wrapper hash was actually used for a given inference. This runtime audit trail is the main distinction from distribution-only signing, which verifies an artifact before use but does not by itself record per-inference wrapper behavior.

\paragraph{Overhead of TIF and BAH.}
TIF and BAH are lightweight runtime checks applied after a trusted wrapper reference has been established. TIF performs a single canonicalization of the
wrapper followed by one SHA-256 hash, reusing the same operations already required by the conjunctive
gate (Eq. \ref{eq:allowed}). As a result, it introduces no additional asymptotic overhead beyond the gate itself.
BAH appends a constant-size attestation record (\texttt{\{wrapper\_lock, allowed\}}) to the inference
log and incurs negligible cost.

\paragraph{Results.}
When the trusted reference corresponds to the clean wrapper, TIF-BAH blocks modified wrappers before the gated branch executes. In our controlled deployment, this neutralized the conjunctive case while preserving normal model behavior. The accuracy reduction was marginal ($<1\%$ absolute), and the added latency was below $0.5\%$ as reported in Section~\ref{subsec: lattt}.

\paragraph{Trusted reference establishment.}
TIF-BAH assumes a trusted reference digest for the expected wrapper. This reference can be established through a full-bundle signed release manifest, a CI/CD-generated deployment manifest, an enterprise allowlist, or a maintainer-published wrapper hash. Thus, TIF-BAH does not solve the distribution trust problem by itself; instead, it enforces the trusted reference at runtime and records which wrapper version was actually used during inference.

\section{Why this is template-layer specific.}
The mechanism is implemented in wrapper-side code, and we do not claim that conditional middleware logic is unique to AI systems. The AI-specific risk is that LLM/VLM outputs are strongly mediated by textual deployment artifacts that are often treated as auxiliary files rather than integrity-critical components. Wrapper logic can change prompt construction, refusal framing, recommendations, citations, disclaimers, routing hints, and final answer formatting. These changes are often perceived by users as model behavior even though they occur outside the model. Existing verification practices focus mainly on model weights, binaries, prompts, or datasets, leaving this template layer weakly covered.

\section{Fulfillment of Adversary Objectives.}
The conjunctive gate successfully achieves all three adversary objectives.
It maintains \textbf{stealth} by embedding the trigger marker invisibly inside the wrapper and encoding the activation condition through cryptographic digests in metadata, so that the modified files appear normal during quick reviews, automated scans, or version control comparisons.
It ensures persistence under canonicalization-preserving transformations (e.g., whitespace
normalization or comment changes), while intentionally fail-closing under semantic wrapper edits
that alter the canonical form. It guarantees \textbf{exclusivity} since activation depends on secret keyed digests that only the original adversary can reproduce, preventing others from hijacking or replicating the same behavior.
Through these properties the attacker gains a covert, durable, and controlled mechanism for influencing model outputs at the template layer while evading commonly deployed static and provenance-based defenses.

\end{document}